\documentclass{article}

    \PassOptionsToPackage{numbers, compress}{natbib}

\usepackage[preprint]{neurips_2026}
\usepackage{makecell}

\usepackage[utf8]{inputenc} 
\usepackage[T1]{fontenc}    
\usepackage{hyperref}       
\usepackage{url}            
\usepackage{booktabs}       
\usepackage{amsfonts}       
\usepackage{nicefrac}       
\usepackage{microtype}      
\usepackage{xcolor}
\usepackage{amsmath}
\usepackage{graphicx}
\usepackage{multirow}

\usepackage{longtable}   
\usepackage{array}       
\title{Assay-Aware BindingDB: Curating Experimental
Context for Binding Affinity Prediction}

\author{%
  Ming-Hsiu Wu \textsuperscript{1} \quad Xuejiao Shirley Guo \textsuperscript{3} \quad Bingsong Zeng \textsuperscript{2} \\
  \vspace{0.2cm}
  \textbf{Ziqian Xie \textsuperscript{1} \quad Shuiwang Ji \textsuperscript{3} \quad 
  Wenshe Liu \textsuperscript{3} \quad
  Cui Tao \textsuperscript{4} \quad
  Degui Zhi \textsuperscript{1}} \\[1ex]
  \textsuperscript{1}The University of Texas Health Science Center at Houston \\
  \textsuperscript{2}The University of Texas MD Anderson Cancer Center \\
  \textsuperscript{3}Texas A\&M University \\
  \textsuperscript{4}Mayo Clinic\\[1ex] 
  \texttt{\{ming.hsiu.wu,bingsong.zeng,ziqian.xie,degui.zhi\}@uth.tmc.edu} \\
  \texttt{wliu@chem.tamu.edu}, \texttt{\{xjguo6,sji\}@tamu.edu}\\
  \texttt{Tao.Cui@mayo.edu}
}

\begin{document}

\maketitle

\begin{abstract}
Protein--ligand binding affinity prediction is fundamental to
computational drug discovery, yet modern AI-driven models are limited
by pervasive heterogeneity in their training data: bioactivity values
are aggregated across diverse assay types and experimental conditions
without accounting for protocol-level differences, introducing
systematic noise. Existing harmonization approaches either discard
assay-level metadata or collapse it into coarse categorical
distinctions, leaving rich contextual signal unused. We address this
gap with two contributions. First, we introduce Assay-Aware
BindingDB, an assay-metadata augmentation of 74{,}425 BindingDB
protein--ligand pairs spanning four widely used assay types (ITC,
SPR, RBA, and FPA), with structured metadata extracted from the
linked primary literature by a two-stage agentic framework that
decouples contextual evidence extraction from ontology-conditioned
JSON synthesis. Evaluated against domain-expert references, the
framework achieves human-curator-comparable fidelity (presence
F1~$\geq 0.947$, semantic content accuracy~$\geq 0.913$ across all
four assay types). Second, we propose a context-conditioned affinity
prediction model that injects an assay-context embedding into the
Boltz-2 affinity module. On a paper-level held-out split designed to
prevent shortcut learning, a single context-conditioned model reduces
combined-regime MSE from $1.27$ to $1.19$ and raises Pearson
correlation from $0.64$ to $0.67$, with significant gains on SPR and FPA; ITC---the only label- and immobilization-free assay in
the set---shows no improvement, consistent with its design removing
the protocol artifacts the metadata captures. Together, these results
support the central hypothesis that systematically curated assay
metadata is informative signal---not irreducible noise---for
protein--ligand affinity prediction wherever protocol-level variation
drives heterogeneity in reported readouts.
\end{abstract}

\section{Introduction}
Protein–ligand binding affinity prediction is a fundamental problem in computational drug discovery, as affinity governs both the potency and selectivity of candidate therapeutics~\cite{spassov2024binding}. Modern AI-driven affinity prediction models are data-hungry, making large-scale chemogenomic databases such as BindingDB~\cite{liu2007bindingdb, liu2025bindingdb} and ChEMBL~\cite{10.1093/nar/gkad1004, gaulton2012chembl} indispensable training resources. However, the predictive reliability of these models is fundamentally limited by pervasive heterogeneity in the underlying bioactivity data~\cite{kalliokoski2013comparability,kramer2012experimental,landrum2024combining}. This heterogeneity arises primarily from the diversity of binding affinity measures: experimental affinities may be reported as $K_d$, $K_i$, $\mathrm{IC}_{50}$, or $\mathrm{EC}_{50}$, with the choice of measure mainly determined by the assay employed. Compounding this issue, critical experimental conditions—including assay format, pH, temperature, protein concentration, and buffer composition—vary across studies, further aggravating cross-dataset inconsistency. Naively aggregating such data without accounting for these differences introduces systematic error into trained models~\cite{landrum2024combining}. Harmonizing these heterogeneous data sources thus remains a central challenge for training reliable models.

To address this metadata gap, structured curation of assay conditions at scale is essential—yet existing approaches fall short. Manual annotation is prohibitively labor-intensive given the exponential growth of biomedical literature, while rule-based NLP methods lack the flexibility to parse the linguistic variability of experimental descriptions~\cite{machi2023ospar,vaucher2020automated}. Recent advances in autonomous AI agents~\cite{fang2025comprehensive,masterman2024landscape,ferrag2025llm}—systems capable of multi-step reasoning, iterative self-correction, and dynamic tool use—offer a transformative opportunity to harmonize bioactivity metadata retrospectively from primary literature at scale. 

\begin{figure}[htb!]
\centering
\includegraphics[width=1\textwidth]{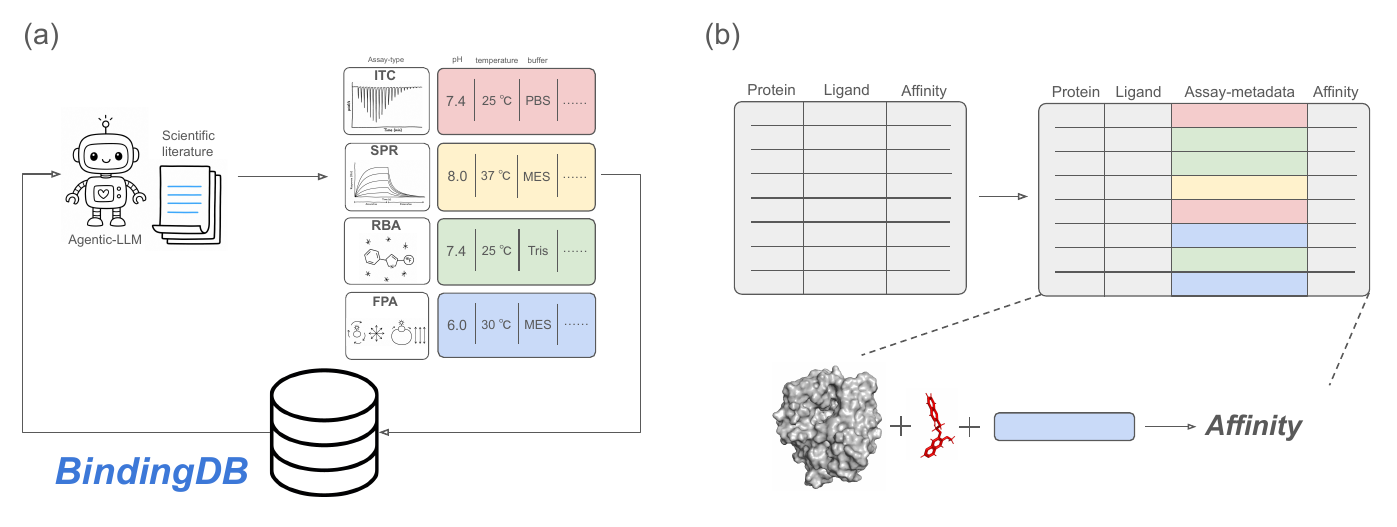}
\caption{Overview. (a) An agentic LLM framework extracts assay metadata from the primary literature linked to BindingDB entries across four assay types (ITC, SPR, RBA, FPA), yielding Assay-Aware BindingDB. (b) The curated metadata augments each (protein, ligand, affinity) record and is injected into the affinity prediction model, conditioning affinity prediction on the experimental protocol.}
\label{fig:overview}
\end{figure}

Building on this insight, we argue that binding affinity prediction can be improved by treating assay context not as discardable noise, but as informative signal. Systematically recovered experimental metadata can be incorporated directly into affinity prediction models, enabling them to account for systematic variation arising from differing experimental conditions. This motivates a unified framework that integrates large-scale automated metadata curation with context-aware binding affinity prediction (Figure~\ref{fig:overview}). 

\paragraph{Our contributions are:}
\begin{itemize}
    \item Assay-Aware BindingDB, a large-scale augmentation of BindingDB entries with structured assay metadata—including assay type, assay format, and experimental conditions—systematically extracted from the linked primary literature.
    \item A context-aware affinity prediction model that injects curated assay metadata into the prediction pipeline, allowing the model to explicitly account for measurement heterogeneity arising from differing experimental protocols.
    \item We demonstrate that incorporating assay context as structured input yields improvements in binding affinity prediction accuracy, validating the hypothesis that assay metadata is informative signal rather than irreducible noise.
\end{itemize}

\section{Related Works}
\subsection{Diverse Binding Affinity Measurement Harmonization}
Several approaches have sought to place heterogeneous bioactivity measurements on a common scale. ChEMBL~\cite{gaulton2012chembl, Mendez2019} introduced the pChEMBL value --- defined as $-\log_{10}(\text{molar activity})$ for standardized dose-response endpoints ($\mathrm{IC}_{50}$, $\mathrm{EC}_{50}$, $K_i$, $K_d$, Potency, $\mathrm{ED}_{50}$) --- which enables convenient ranking and thresholding of compound potency but treats all qualifying measurement types as interchangeable without accounting for experimental conditions, or detection methodology. Tang et al.~\cite{tang2014making} proposed the KIBA score, a parametric integration of $\mathrm{IC}_{50}$, $K_i$, and $K_d$ via Cheng-Prusoff-inspired~\cite{yung1973relationship} adjustments that improved drug--target interaction classification over any single bioactivity type; however, KIBA is restricted to kinase targets, harmonizes only at the level of bioactivity type without capturing assay protocol details, and applies a global two-parameter correction that assumes a fixed $\mathrm{IC}_{50}$--$K_i$/$K_d$ relationship across all systems. Tanoli et al.~\cite{tanoli2020interactive} introduced DTP scoring, which maps bioactivity values onto a $[0, 1]$ potency scale using protein-family-specific cutoffs (from IDG/TCRD) and a binary assay-format factor (biochemical vs.\ cell-based), parameterizing a piecewise linear transformation with fixed thresholds and aggregating multiple measurements per compound--target pair by median. While DTP improves on pChEMBL's uniform $-\log_{10}$ transform by introducing family- and format-level awareness, it still treats all biochemical assays identically (e.g., surface plasmon resonance, fluorescence polarization, and radioligand binding share the same factor) and relies on hand-crafted rules. Across all three approaches, the core limitation is the same: assay-level protocol metadata --- the detection technology, binding model, and experimental conditions that shape the reported value --- is either discarded or collapsed into coarse categorical distinctions.

\subsection{Agent-Based Information Extraction}
While databases like ChEMBL and BindingDB contain millions of curated records, they represent only a small fraction of historical and contemporary medicinal chemistry data. The vast majority of protein--ligand bioactivity measurements remain trapped within the unstructured text, tables, and supplementary materials of scientific literature and pharmaceutical patents. Recent multi-agent LLM systems have begun to address this bottleneck~\cite{inoue2025drugagent,chen2025multi}. Yan et al.~\cite{yan2025biominer} introduced \textsc{BioMiner}, a multi-modal agentic system that decomposes bioactivity mining from journal articles into four specialized agents for document parsing, chemical structure resolution, bioactivity measurement extraction, and cross-modal integration, explicitly separating semantic reasoning---handled by a domain-tuned multi-modal LLM---from deterministic chemical symbolic construction via RDKit~\cite{rdkit} and OPSIN~\cite{lowe2011chemical} to enable reliable resolution of complex Markush representations. Concurrently, Shepard et al.~\cite{shepard2026harvest} presented \textsc{Harvest}, a five-agent sequential pipeline targeting USPTO patent archives, which chains three LLM-based semantic agents---target identification, activity extraction conditioned on upstream target context, and compound alias resolution---with two deterministic resolution agents for chemical structure standardization and protein mapping to UniProt identifiers. Both systems demonstrate that principled agent decomposition can scale bioactivity extraction to tens of thousands of documents with near-human fidelity. However, neither system captures or encodes the assay-level context---detection technology, binding model, and experimental conditions---that governs cross-protocol comparability of the extracted values. Our work aims to fill this gap.

\section{Assay-Aware BindingDB}

\subsection{Two-Stage Agentic Extraction Framework}
\label{sec:framework}

\paragraph{Problem Formulation}

We decompose the heterogeneous source $\mathcal{D}$ into two functionally
distinct components: a BindingDB record $r$ that provides a brief description of the target assay, and the associated scientific
literature $\mathcal{L}$ containing the full experimental details. That is,
$\mathcal{D} = (r,\, \mathcal{L})$. The record $r$ acts as a \emph{query}
anchoring the extraction task, while $\mathcal{L}$ is the corpus to be
mined. Together with a predefined assay schema ontology $\mathcal{O}$, the
goal is to construct a structured ontology-compliant record $\mathcal{C}_s$
\begin{equation}
    \Phi : \mathcal{L} \mid r,\, \mathcal{O}
    \;\longrightarrow\; \mathcal{C}_s,
    \qquad \text{such that} \quad \mathcal{C}_s \models \mathcal{O},
    \label{eq:phi}
\end{equation}
where $r$ and $\mathcal{O}$ appear as conditioning inputs (prompts) rather
than as data to be transformed, and $\mathcal{C}_s \models \mathcal{O}$
denotes schema satisfaction. We factorize $\Phi$ through an intermediate
semantic context $\mathcal{C}_e$ (Figure~\ref{fig:two_stage_agentic_extraction_framework}):
\begin{equation}
\mathcal{C}_e \;=\; f_{\text{ext}}(\mathcal{L} \mid r), \qquad \Phi(\mathcal{L} \mid r, \mathcal{O}) \;=\; f_{\text{struct}}(\mathcal{C}_e \mid \mathcal{O}).
\label{eq:factorization}
\end{equation}
Throughout, we use the bar ``$\mid$'' to denote conditioning: the argument
to its left is transformed by the function, whereas the argument to its
right enters as a prompt that shapes the function's behavior without being
itself rewritten. In what follows, we instantiate $f_\text{ext}$ and $f_\text{struct}$ as two LLM agents that respectively realize the contextual extraction and ontology-conditioned synthesis stages of Eq.~(2).

\paragraph{Agent 1 --- Contextual evidence extractor.}
\label{subsec:agent1}

Agent~1 is prompted with the BindingDB record $r$---a concise
natural-language description of the assay of interest (e.g., \emph{``Binding
of inhibitors and substrates to immobilized AcrB''})---and uses it as a
query to search the associated literature $\mathcal{L}$ (the main paper, with supplementary information and referenced papers when accessible) for supporting
evidence:
\begin{equation}
    f_{\text{ext}} : \mathcal{L} \mid r
    \;\longrightarrow\; \mathcal{C}_e,
    \qquad
    \mathcal{C}_e = \{c_1, c_2, \ldots, c_n\}.
    \label{eq:fext}
\end{equation}
Each context unit $c_i$ is a localized evidence span (a text segment, figure, table ,or caption) drawn from $\mathcal{L}$, annotated with a provenance tag (e.g.,
\textsc{[experimental procedures]}, \textsc{[figure]},
\textsc{[supplementary]}). Because $r$ only
enters as a conditioning prompt, $f_{\text{ext}}$ does not attempt to
rewrite or transform it; it uses $r$ solely to focus attention on the
assay-relevant portions of $\mathcal{L}$.

\paragraph{Agent 2 --- JSON Synthesizer.}
\label{subsec:agent2}

Agent~2 is prompted with the predefined assay schema ontology
$\mathcal{O}$---a machine-readable specification of admissible fields,
types, descriptions, and enumerations---and maps the extracted semantic
context $\mathcal{C}_e$ to an ontology-compliant record:
\begin{equation}
    f_{\text{struct}} : \mathcal{C}_e \mid \mathcal{O}
    \;\longrightarrow\; \mathcal{C}_s,
    \qquad \mathcal{C}_s \models \mathcal{O}.
    \label{eq:fstruct}
\end{equation}
Here $\mathcal{O}$ is not consumed as data but conditions the decoding
behavior of $f_{\text{struct}}$, constraining the output space so that
every key--value pair in $\mathcal{C}_s$ conforms to a declared field in
$\mathcal{O}$ and is grounded in one or more context units from
$\mathcal{C}_e$. 

Both agents are instantiated with Qwen3.5-27B~\cite{qwen3, qwen3.5} in non-thinking mode under greedy decoding ($T = 0$). Agent 1 reads rasterized page images of the main paper, supplementary PDFs, and accessible cited references; Agent 2 consumes the resulting spans in text-only mode under ontology-conditioned prompting. When Agent 1 indicates the protocol is deferred to a cited paper, the citation is resolved against a MinerU markdown~\cite{niu2025mineru2} rendering of the source reference list and the extraction is recursed on the cited paper up to a fixed depth. Extraction is run once per unique (\texttt{PMID}, \texttt{DESCRIPTION}) and shared across all pairs in that group. See Appendix~\ref{app:implementation} for full details.

\subsection{Extraction Evaluation}
\label{sec:eval}
\paragraph{Protocol.}
We evaluate the predicted record $\mathcal{C}_s$ against a human-curated reference $\mathcal{C}_s^{*}$ at the leaf-field level, under a typed, multi-metric protocol designed to disentangle genuine extraction errors from superficial format, unit, and paraphrase variation. The leaf fields of $\mathcal{O}$ are partitioned by value space into four types---\emph{categorical} (e.g., instrument manufacturer, sensor-chip type, immobilization strategy), \emph{numeric scalar} (pH, temperature, analyte flow rate), \emph{numeric range} (immobilization density, association/dissociation time), and \emph{free text} (buffer composition, analyte concentration ranges, reference description)---and each type is paired with a dedicated comparator. Categorical fields are scored by normalized exact match after Unicode folding, case folding, and whitespace collapsing. Numeric scalars are compared only after unit harmonization: for each dimension $D \in \{\text{time}, \text{temperature}, \text{flow rate}, \text{concentration}\}$ we define a canonical unit $u_D^{*}$ (respectively $\text{s}$, $^{\circ}\text{C}$, $\mu\text{L}/\text{min}$, $\text{nM}$) and a map
\begin{equation}
\phi_D : (v, u) \;\longmapsto\; v^{*} \in \mathbb{R},
\end{equation}
that parses a value--unit pair $(v, u)$ and converts it to the canonical
scale. The unit $u$ may be specified in one of three ways: embedded within
the value string (e.g., \texttt{"3 min"}), provided in a sibling field
(e.g., \texttt{flow\_rate.value} paired with \texttt{flow\_rate.unit}), or
absent from the record, in which case $\varphi_D$ falls back to the
ontology default $u^*_D$. Conversion is linear across all three dimensions.
Two scalars are then deemed equal when their canonical forms agree within
a 2\% relative tolerance, so that $(3, \text{min})$ and $(180, \text{s})$
are correctly identified as equivalent. Numeric ranges are compared endpoint-wise under the same tolerance after range-separator normalization. Free-text fields are scored by an LLM (Qwen3.5-27B) judge
\begin{equation}
\mathcal{J}(p, g \mid k, d) \longrightarrow \{0.0,\ 0.5,\ 1.0\},
\end{equation}
conditioned on the field name $k$ and its ontology description $d \in \mathcal{O}$ so that decisions are grounded in field-specific semantics, with $1.0$ denoting factual equivalence, $0.5$ denoting a strict informational superset/subset relation (a common and benign outcome when concise paraphrases drop non-essential detail), and $0.0$ denoting factual disagreement or hallucination; each comparison is averaged over $K = 3$ independent judge samples to reduce variance.

We decouple null-handling from content quality by reporting four orthogonal axes: (i) \emph{presence}---precision, recall, and F1 over the extractor's null vs.\ non-null decisions, which penalizes both missed fields (FN) and hallucinated values (FP); (ii) \emph{content accuracy} on TP rows, reported at two strictness levels---normalized (type-, unit-, and format-aware) equality, and semantic equivalence under $\mathcal{J}$---so that the normalized-to-semantic gap quantifies paraphrase tolerance independently of genuine content errors; (iii) \emph{per-type semantic content accuracy} on TP rows, reported separately for each of the four field types under its most permissive comparator (semantic equivalence under $\mathcal{J}$ for free text, normalized equality for categorical, numeric scalar, and numeric range), so that extraction bottlenecks tied to specific value spaces---e.g., free-text paraphrase fidelity vs.\ numeric unit handling---are not washed out by aggregation; and (iv) \emph{overall field accuracy}, which credits the extractor for both correctly predicted nulls (TN) and correctly extracted values (TP) at each of the two strictness levels, providing a single composite score that reflects end-to-end record-construction quality. We report micro averages pooled across record--field pairs within each assay type, enabling per-assay-type comparison of extraction quality.

\paragraph{Results.}
\label{extraction_result}
We instantiate the framework of Section~\ref{sec:framework} on four widely used assay types reported in BindingDB: isothermal titration calorimetry (ITC), surface plasmon resonance (SPR), radioligand binding assay (RBA), and fluorescence polarization anisotropy (FPA). For each BindingDB entry, the assay type is assigned by keyword matching against their \texttt{DESCRIPTION} field, and a type-specific schema ontology $\mathcal{O}$ is used to condition Agent~2. The full schema ontologies $\mathcal{O}$ for these four assay types are provided in Appendix~\ref{app:schemas}. To evaluate extraction quality, we uniformly sample 30 distinct papers per assay type (120 papers in total); since a single paper may contribute multiple records, the resulting evaluation set covers a larger number of extracted records $\mathcal{C}_s$, each of which is compared against a domain-expert-curated reference $\mathcal{C}_s^*$ under the protocol of Section~\ref{sec:eval}. Results across all four assay types are reported in Table~\ref{tab:extraction_results}.

\begin{table}[h]
\centering
\caption{Four-axis evaluation of the agentic extraction framework across four assay types (ITC, SPR, RBA, FPA). Rows report presence (precision/recall/F1 over null vs.\ non-null decisions), content accuracy on TP rows at two strictness levels, per-type semantic accuracy on TP rows, and overall field accuracy with TN credited as correct. All numbers are micro averages within each assay type.}
\label{tab:extraction_results}
\small
\begin{tabular}{lcccc}
\toprule
 & ITC & SPR & RBA & FPA \\
\midrule
\multicolumn{5}{l}{\textit{Presence (null vs.\ non-null)}} \\
\quad Precision & 0.994 & 0.958 & 0.942 & 0.964 \\
\quad Recall    & 0.980 & 0.988 & 0.951 & 0.983 \\
\quad F1        & 0.987 & 0.973 & 0.947 & 0.974 \\
\midrule
\multicolumn{5}{l}{\textit{Content accuracy (TP rows)}} \\
\quad Normalized & 0.907 & 0.785 & 0.957 & 0.955 \\
\quad Semantic   & 0.954 & 0.913 & 0.994 & 0.977 \\
\midrule
\multicolumn{5}{l}{\textit{Per-type semantic content accuracy (TP rows)}} \\
\quad Categorical    & 1.000 & 0.835 & 0.988 & 0.965 \\
\quad Numeric scalar & 0.936 & 0.885 & 1.000 & 0.992 \\
\quad Numeric range  & 0.890 & 0.933 & 1.000 & 0.870 \\
\quad Free text      & 1.000 & 1.000 & 1.000 & 1.000 \\
\midrule
\multicolumn{5}{l}{\textit{Overall field accuracy (TN credited as correct)}} \\
\quad Normalized & 0.919 & 0.794 & 0.911 & 0.936 \\
\quad Semantic   & 0.951 & 0.891 & 0.932 & 0.950 \\
\bottomrule
\end{tabular}
\end{table}

\paragraph{Overall fidelity.} Across all four assay types, the framework
reliably both detects which assay-condition fields are present in the source
literature and extracts semantically equivalent values when they are.
Presence F1 ranges from 0.947 (RBA) to 0.987 (ITC), and semantic content
accuracy from 0.913 (SPR) to 0.994 (RBA). The composite overall field
accuracy, which jointly credits correct null predictions and correct
extractions, exceeds 0.89 for every assay type, indicating that the
two-stage agentic framework produces structured assay metadata at
human-curator-comparable fidelity.

\paragraph{Normalized vs.\ semantic.} The normalized-to-semantic gap is largest for SPR ($+0.128$), reflecting the prevalence of fields whose information content is preserved across substantially different surface forms. Analyte concentration ranges, for instance, may be reported as explicit endpoints (\emph{``3.12--25\,\textmu M''}) or as a top concentration plus a dilution scheme (\emph{``25\,\textmu M highest, 4 concentrations, dilution factor 0.5''})---semantically equivalent specifications that normalized equality cannot reconcile but $\mathcal{J}$ recognizes as factually consistent. Buffer composition exhibits analogous variability, with identical formulations rendered in different component orderings, abbreviations, and concentration notations. Only $\mathcal{J}$ accommodates this surface-form heterogeneity.

\paragraph{Field-type breakdown.} At the leaf-field-type level, free-text fields achieve $1.000$ semantic
accuracy across all four assays, confirming that Agent 2 reliably captures the substantive content. Categorical fields are extracted with $\geq 0.96$ semantic
accuracy except for SPR ($0.835$), where the larger and more heterogeneous vocabulary of biosensor instruments, chip chemistries, and immobilization protocols leaves room for residual misclassification. Part of this residual
error is itself irreducible, arising from ambiguous or incomplete descriptions in the source literature for fields such as \texttt{data\_analysis.fitting\_model}, which leave the ground-truth assignment underdetermined and thus place a ceiling on attainable extraction accuracy. Numeric fields are extracted with $\geq 0.87$ semantic accuracy across the board, with SPR again the most challenging case in numeric scalar and FPA slightly weaker on numeric ranges. Taken together, these results validate the central design choice of Section 3.1—agentic context extraction followed by ontology-conditioned synthesis—and establish that the recovered metadata is of sufficient quality to serve as informative signal for the context-conditioned affinity prediction model of Section 4.

\subsection{Dataset Construction}
\label{sec:assay-aware-bindingdb}
Having established that the two-stage extraction framework produces metadata at human-curator-comparable fidelity, we apply it to the subset of BindingDB entries matching our four target assay types (ITC, SPR, RBA, FPA) to construct \textbf{Assay-Aware BindingDB}. Construction proceeds in three stages. First, we identify candidate entries by keyword matching the assay-type descriptors against BindingDB's \texttt{DESCRIPTION} field, yielding 113{,}311 candidate protein--ligand pairs. Second, for each candidate we attempt to retrieve the associated primary literature; after removing entries whose source articles are not accessible, 5{,}991 unique papers remain. Third, we apply the two-stage agentic extractor of Section~\ref{sec:framework} to every retrieved paper and retain only those records whose extracted metadata satisfies the assay-type ontology $\mathcal{O}$. After this final filter, \textbf{74{,}425 protein--ligand pairs} constitute Assay-Aware BindingDB, each paired with a structured, ontology-compliant assay-context record suitable for downstream context-conditioned modeling.

Table~\ref{tab:dataset-composition} reports the per-assay-type composition. The dataset spans 1{,}662 unique protein targets and 46{,}218 unique compounds across four complementary slices of the bioactivity landscape: ITC (1{,}898 pairs) and SPR (4{,}253 pairs) report exclusively $K_d$ values, consistent with their thermodynamic and kinetic readouts; RBA contributes the largest single cohort (49{,}291 pairs), dominated by $K_i$ (34{,}516) and $\mathrm{IC}_{50}$ (14{,}853) values typical of competitive radioligand-displacement experiments; and FPA (18{,}983 pairs) contributes a mixed-measure cohort spanning all three affinity types. This curated, context-annotated resource forms the basis of the affinity-prediction experiments in Section~\ref{sec:context-conditioned-prediction}. Attrition between candidate identification and the final dataset arises predominantly from constraints on source-text availability and processing rather than extraction-quality failures; we defer a detailed discussion to the Limitations section.

\begin{table}[t]
  \caption{Composition of Assay-Aware BindingDB. For each assay type, we report the number of candidate BindingDB entries matching the assay-type keywords, the number of papers retrieved, the retained entries (protein–ligand pairs), the number of unique targets and compounds, and the counts of reported Kd, Ki, and IC50 values. Unique target, compound, and affinity-measure counts are computed over the final retained pairs.}
  \label{tab:dataset-composition}
  \centering
  \small
  \begin{tabular*}{\textwidth}{@{\extracolsep{\fill}} c|c|c|c|ccccc@{}}
    \toprule
    \makecell{Assay\\type} & \makecell{Candidates\\(pair)} & \makecell{Lit.\\retrieved} & \makecell{Retained\\(pair)} & Targets & Compounds & $K_d$ & $K_i$ & IC$_{50}$ \\
    \midrule
    ITC & 2,157  & 346   & 1,898  & 272   & 1,382  & 1,898 & -      & -      \\
    SPR & 4,883  & 590   & 4,253  & 424   & 3,351  & 4,253 & -      & -      \\
    RBA & 77,051 & 4,343 & 49,291 & 593   & 30,136 & -     & 34,516 & 14,853 \\
    FPA & 29,220 & 712   & 18,983 & 373   & 11,349 & 2,385 & 4,909  & 11,689 \\
    \midrule
    Total & 113,311 & 5,991 & 74,425 & 1,662 & 46,218 & 8,536 & 39,425 & 26,542 \\
    \bottomrule
  \end{tabular*}
\end{table}

\section{Context-conditioned Binding Affinity Prediction}
\label{sec:context-conditioned-prediction}

\paragraph{Assay-context encoding.}
Given an ontology-compliant record $\mathcal{C}_s$ produced by the extraction pipeline of Section~\ref{sec:framework}, we augment it with two additional fields---the assay type (e.g., ITC, SPR, FPA, RBA) and the reported binding-affinity measure (e.g., $K_d$, $K_i$, $\text{IC}_{50}$)---and render the resulting
record as a JSON string $\tau(\mathcal{C}_s)$. This string is encoded
with the pretrained Qwen3-Embedding-8B encoder $\phi_{\text{ctx}}$:
\begin{equation}
  \mathbf{e}_{\text{ctx}} \;=\; \phi_{\text{ctx}}\!\bigl(\tau(\mathcal{C}_s)\bigr)
  \;\in\; \mathbb{R}^{d_{\text{ctx}}},
  \qquad d_{\text{ctx}} = 4096.
\end{equation}
Using a general-purpose text encoder allows us to represent arbitrary
assay fields as a single dense vector without hand-designing a separate
embedding for each categorical attribute, and naturally accommodates
free-text descriptors that resist one-hot
encoding.

\paragraph{Injection into the Boltz-2 affinity module.}
We adopt Boltz-2~\cite{passaro2025boltz} as our base structural model and modify its affinity module to condition on $\mathbf{e}_{\text{ctx}}$. The original module mean-pools the refined pair representation $z$ over the protein--ligand interaction mask to obtain a trunk summary $\mathbf{g}$, which is then passed
through an MLP and two heads predicting a binding likelihood and a
numerical affinity value. We retain only the affinity regression head
and intercept immediately after the pooling step, concatenating the
assay-context embedding to $\mathbf{g}$ before the downstream MLP:
\begin{align}
  \mathbf{g} &= \operatorname{MeanPooling}\!\bigl(z,\ \text{mask} =
    M_{\text{pl}} + M_{\text{il}}\odot(1 - \text{Id})\bigr), \\
  \tilde{\mathbf{g}} &= \bigl[\,\mathbf{g}\;\|\;\mathbf{e}_{\text{ctx}}\,\bigr], \\
  \hat{y}_{\text{aff}} &= \operatorname{MLP}_{\text{aff}}(\tilde{\mathbf{g}}).
\end{align}
where $[\,\cdot\;\|\;\cdot\,]$ denotes vector concatenation,
$M_{\text{pl}}$ and $M_{\text{il}}$ are the protein--ligand and
intra-ligand masks, and $\operatorname{MLP}_{\text{aff}}$ is a five-layer MLP --- four ReLU-activated linear layers followed by a final linear projection --- inherited from Boltz-2's post-pooling pathway and affinity regression head, with its first linear layer widened from $d_g$ to $d_g + d_{\text{ctx}}$.

\paragraph{Training.}
Given the limited size of our training set, we train only $\operatorname{MLP}_{\text{aff}}$ from scratch on the curated Assay-Aware BindingDB while keeping the rest of the publicly released Boltz-2 weights frozen, allowing the model to focus on learning the assay-context conditioning. The text encoder $\phi_{\text{ctx}}$ is likewise kept frozen throughout training to preserve its pretrained semantic space; only $\operatorname{MLP}_{\text{aff}}$ (including its widened first linear layer) receives gradient updates.

\subsection{Experimental Results}
\label{sec:results}

\paragraph{Splitting strategy.}
We split Assay-Aware BindingDB at the paper level rather than at the
protein--ligand pair level to prevent shortcut learning through the
assay-context embedding. Pairs reported within a single paper typically span
a narrow range of binding affinities, since they share the same assay
protocol, target, and chemotype series. Under a pair-level split, the
assay-context embedding $e_{\text{ctx}}$ would act as a near-unique identifier
of the source paper: rather than learning how assay conditions modulate
affinity, the model could exploit $e_{\text{ctx}}$ as a shortcut to recognize
the paper at test time and predict any value within its narrow affinity
range, inflating metrics without genuine context-aware generalization.
Paper-level splitting eliminates this shortcut by ensuring that no
assay-context embedding seen during training recurs at test time, so any
improvement attributable to $e_{\text{ctx}}$ must reflect generalization
across assay protocols rather than memorization of paper identity.

\paragraph{Experimental setup.}
We evaluate three configurations on the paper-level held-out split:
(i)~the full model with assay-context embedding (\emph{w/});
(ii)~an ablation in which $e_{\text{ctx}}$ is removed and predictions are made
from the structural trunk summary $g$ alone (\emph{w/o}); and
(iii)~a context-only variant (\emph{ctx-only}) in which the structural
representation is ablated and predictions are made from $e_{\text{ctx}}$
alone. We report mean squared error (MSE) on $p$-affinity values
($-\log_{10}$ affinity in M), Pearson correlation ($R_p$), and concordance
index (c-index), each as mean~$\pm$~std over 10 independent paper-level
train/val/test splits. We further exclude pairs whose reported affinity is censored (i.e., expressed as a $>$ or $<$ inequality rather than a point estimate) and pairs whose protein or ligand inputs fail preprocessing by the Boltz-2 structural pipeline or RDKit. We train and evaluate in two regimes: a \emph{combined}
regime, in which the model is trained on the union of all four assay types
and evaluated on the full held-out test set (reported as \emph{All} in
Table~\ref{tab:affinity-results}); and four \emph{per-assay-type} regimes,
in which a separate model is trained and evaluated on each assay type in
isolation (ITC, SPR, FPA, RBA). The per-assay-type regimes isolate the
contribution of assay-context conditioning within each protocol family,
while the combined regime tests whether $e_{\text{ctx}}$ enables a single
model to handle protocol heterogeneity across the dataset.

\paragraph{Aggregate results.}
Across the combined regime (65{,}169 pairs, all four assay types
pooled), conditioning on assay context reduces MSE from $1.27$ to
$1.19$, raises $R_p$ from $0.64$ to $0.67$, and improves c-index
from $0.73$ to $0.74$. A paired $t$-test across the 10 splits finds
all three improvements significant ($p < 0.001$ on MSE, $R_p$,
and c-index). The ctx-only variant collapses on every metric (MSE
$1.77$, $R_p$ $0.45$, c-index $0.65$), confirming that
$e_{\text{ctx}}$ supplies signal complementary to---not a shortcut
around---the structural representation, and that paper-level splitting
effectively removes the trivial-identifier failure mode discussed
above.

\paragraph{Per-assay-type breakdown.}
Within each per-assay-type regime, the gains are most pronounced on
SPR and FPA. SPR shows the largest single-cohort
improvement: MSE drops by 15\% ($2.47 \rightarrow 2.11$), $R_p$ rises
by $+0.09$ ($0.50 \rightarrow 0.59$), and c-index by $+0.03$
($0.67 \rightarrow 0.71$), with all three gains significant at
$p < 0.01$. FPA shows similarly
broad and significant improvements across MSE
($1.06 \rightarrow 0.94$, $-11\%$), $R_p$ ($+0.04$), and c-index
($+0.01$), again with $p < 0.01$ on all three. RBA---the largest cohort, where the structural baseline is already
strongest---shows smaller but consistent gains: MSE $1.17 \rightarrow
1.15$ (not significant, $p = 0.14$), $R_p$ $+0.02$ ($p < 0.05$), and
c-index $+0.01$ ($p < 0.05$).

ITC is the lone exception: under context conditioning,
MSE rises from $1.35$ to $1.47$ ($p < 0.05$), $R_p$ drops from $0.49$
to $0.41$ ($p < 0.1$), and c-index regresses by $0.03$ without
reaching significance. We argue this regression is consistent with,
rather than contrary to, the central hypothesis of the paper. ITC is
widely regarded as the gold-standard biophysical method for
binding-affinity measurement precisely because it is label-free and
immobilization-free: both partners are free in solution, and $K_d$ is
recovered as a direct thermodynamic quantity without recourse to an
exogenous reporter~\cite{vega2015unified,krainer2015single,bastos2023isothermal}. Each absent component is itself a
protocol-dependent source of variation that the schema is designed to
capture in the other three assays---tracer $K_d$ uncertainty
propagating through Cheng--Prusoff in RBA and FPA, surface-density and
mass-transport effects in SPR, fluorescence interference and label
perturbation in FPA---but is, by construction, absent from ITC. The
recoverable protocol metadata for ITC therefore modulates the reported
$K_d$ far less than its counterparts modulate the readouts of the
other three assays, and when the affinity-modulating signal in
$e_{\text{ctx}}$ is weak relative to the parameters it adds to the
head, conditioning on it can introduce noise rather than information.

\paragraph{On the value of curated assay context.}
Two observations support the claim that the curated assay metadata is
a meaningful source of these gains. First, both the \emph{w/} and
\emph{w/o} configurations are evaluated on top of Boltz-2 weights
pretrained on BindingDB, so any baseline performance attributable to
seeing protein--ligand pairs during pretraining is shared by both
configurations and cancels out of the comparison. The improvement of
\emph{w/} over \emph{w/o} therefore isolates the contribution of the
assay-context embedding $e_{\text{ctx}}$, independent of pretraining
exposure. Second, the gains delivered by $e_{\text{ctx}}$---significant in the
combined regime and in three of the four per-assay-type regimes
(SPR, FPA, RBA)---indicate that the assay metadata supplies
information beyond what the structural trunk already encodes. Taken together, the dataset and the
context-conditioned head support the central hypothesis of this work:
systematically curated assay metadata is genuine, non-trivial signal
for protein--ligand affinity prediction wherever protocol-level
variation drives heterogeneity in reported affinity values, and
treating it as such is a productive lever for closing the gap between
heterogeneous bioactivity data and reliable affinity prediction.

\begin{table}[!htbp]
\centering
\caption{Context-conditioned affinity prediction on the paper-level
held-out split of Assay-Aware BindingDB. The \emph{All} row reports the
combined regime (single model trained on all four assay types pooled);
the ITC, SPR, FPA, and RBA rows report per-assay-type regimes (separate
model trained on each assay type in isolation). \emph{w/}: full model
with $e_{\text{ctx}}$; \emph{w/o}: $e_{\text{ctx}}$ ablated;
\emph{ctx-only}: structural trunk ablated. Reported as mean~$\pm$~std
over 10 independent paper-level train/val/test splits. Best result per
metric per cohort in \textbf{bold}.}
\label{tab:affinity-results}
\setlength{\tabcolsep}{4pt}
\renewcommand{\arraystretch}{1.05}
\small
\begin{tabular}{@{}llccc@{}}
\toprule
Cohort & Variant & MSE $\downarrow$ & $R_p$ $\uparrow$ & c-index $\uparrow$ \\
\midrule
\multirow{3}{*}{All (65{,}169)}
 & w/        & $\mathbf{1.19{\pm}0.05}$ & $\mathbf{0.67{\pm}0.02}$ & $\mathbf{0.74{\pm}0.01}$ \\
 & w/o       & $1.27{\pm}0.06$          & $0.64{\pm}0.02$          & $0.73{\pm}0.01$          \\
 & ctx-only  & $1.77{\pm}0.07$          & $0.45{\pm}0.02$          & $0.65{\pm}0.01$          \\
\cmidrule(l){1-5}
\multirow{3}{*}{ITC (1{,}761)}
 & w/        & $1.47{\pm}0.27$          & $0.41{\pm}0.17$          & $0.64{\pm}0.06$          \\
 & w/o       & $\mathbf{1.35{\pm}0.26}$ & $\mathbf{0.49{\pm}0.07}$ & $\mathbf{0.67{\pm}0.03}$ \\
 & ctx-only  & $1.64{\pm}0.38$          & $0.22{\pm}0.15$          & $0.58{\pm}0.04$          \\
\cmidrule(l){1-5}
\multirow{3}{*}{SPR (3{,}650)}
 & w/        & $\mathbf{2.11{\pm}0.27}$ & $\mathbf{0.59{\pm}0.07}$ & $\mathbf{0.71{\pm}0.03}$ \\
 & w/o       & $2.47{\pm}0.14$          & $0.50{\pm}0.10$          & $0.67{\pm}0.04$          \\
 & ctx-only  & $2.83{\pm}0.48$          & $0.39{\pm}0.14$          & $0.63{\pm}0.05$          \\
\cmidrule(l){1-5}
\multirow{3}{*}{RBA (44{,}223)}
 & w/        & $\mathbf{1.15{\pm}0.07}$ & $\mathbf{0.61{\pm}0.02}$ & $\mathbf{0.72{\pm}0.01}$ \\
 & w/o       & $1.17{\pm}0.08$          & $0.59{\pm}0.02$          & $0.71{\pm}0.01$          \\
 & ctx-only  & $1.70{\pm}0.06$          & $0.27{\pm}0.03$          & $0.59{\pm}0.01$          \\
\cmidrule(l){1-5}
\multirow{3}{*}{FPA (15{,}535)}
 & w/        & $\mathbf{0.94{\pm}0.09}$ & $\mathbf{0.73{\pm}0.03}$ & $\mathbf{0.76{\pm}0.01}$ \\
 & w/o       & $1.06{\pm}0.12$          & $0.69{\pm}0.05$          & $0.75{\pm}0.02$          \\
 & ctx-only  & $1.61{\pm}0.22$          & $0.47{\pm}0.07$          & $0.65{\pm}0.02$          \\
\bottomrule
\end{tabular}
\end{table}

\section{Limitations}
\label{sec:limitations}

\paragraph{Metadata extraction coverage.}
While extraction is human-curator-comparable on records it processes (Section~\ref{extraction_result}), many candidate BindingDB entries fail upstream of the extractor: (i) the source article is inaccessible to our retrieval pipeline; (ii) operative protocol details are deferred to cited methods papers or supplementary material we cannot retrieve; or (iii) the paper exceeds the language model's context window or memory budget. A residual fraction reflects under-specified source descriptions---an irreducible ceiling regardless of model capacity. Better retrieval (supplementary material, cited references), longer-context extractors, and targeted re-prompting are natural avenues.

\paragraph{Assay-type coverage.}
We cover four assay types (ITC, SPR, RBA, FPA); others---ELISA, thermal shift, AlphaScreen, and various cell-based functional readouts---fall outside our scope. Extending $\mathcal{O}$ and the extraction prompts to these assays would broaden coverage and test cross-modality generality.

\paragraph{Context encoding and model architecture.}
The structured assay record is serialized to JSON, embedded by a frozen general-purpose text encoder, and concatenated with the Boltz-2 trunk summary before the affinity head. We have not systematically explored alternative context representations (per-field embeddings, ontology-aware encoders, graph representations of buffer and protocol topology, or categorical--text hybrids) or fusion architectures (cross-attention, context-conditioned modulation of earlier trunk layers). These may yield further gains over our concatenation baseline.

\section{Conclusion}
\label{sec:conclusion}

We presented Assay-Aware BindingDB: 74{,}425 BindingDB protein--ligand
pairs augmented with structured assay metadata, extracted from primary
literature by a two-stage agentic framework that attains
human-curator-comparable fidelity across ITC, SPR, RBA, and FPA.
Injecting an assay-context embedding into the Boltz-2 affinity module
reduces combined-regime MSE from $1.27$ to $1.19$ and raises Pearson
correlation from $0.64$ to $0.67$ on a paper-level held-out split,
with significant gains on SPR, FPA, and RBA. ITC---the only label-
and immobilization-free assay---shows no improvement, consistent with
its design removing the very protocol artifacts that the metadata
captures. These results support the
hypothesis that curated assay metadata is informative signal, not
irreducible noise, wherever protocol-level variation drives
heterogeneity in reported affinity values.

\setcitestyle{numbers,square}
\bibliographystyle{plainnat}
\bibliography{reference}

\newpage
\appendix
\renewcommand{\thefigure}{A\arabic{figure}}
\setcounter{figure}{0}
\section{Assay Schema Ontologies}
\label{app:schemas}
This appendix specifies the four type-specific schema ontologies $\mathcal{O}$ used
to condition Agent~2 (Section~3.1) during ontology-compliant JSON synthesis. Each
schema enumerates the leaf fields, their value types, whether they are required by
the schema, and either their permitted enumerated values or representative examples
drawn from the surveyed literature. Field paths use dot notation to indicate
nesting. A dagger ($\dagger$) marks fields the schema declares as required. The
\texttt{type?} suffix denotes a nullable type (i.e., \texttt{null} is permitted in
addition to the listed type).

\subsection{Isothermal Titration Calorimetry (ITC)}
\label{app:schema:itc}
 
\begin{small}
\begin{longtable}{@{}>{\raggedright\arraybackslash}p{0.31\textwidth}>{\raggedright\arraybackslash}p{0.13\textwidth}>{\raggedright\arraybackslash}p{0.50\textwidth}@{}}
\caption{Schema ontology $\mathcal{O}$ for Isothermal Titration Calorimetry (ITC). A dagger ($\dagger$) marks fields required by the schema. \texttt{type?} denotes a nullable type.}\\
\label{tab:schema:itc} \\
\toprule
\textbf{Field path} & \textbf{Type} & \textbf{Description} \\
\midrule
\endfirsthead
\multicolumn{3}{c}{\textit{(continued from previous page)}} \\
\toprule
\textbf{Field path} & \textbf{Type} & \textbf{Description} \\
\midrule
\endhead
\midrule
\multicolumn{3}{r}{\textit{(continued on next page)}} \\
\endfoot
\bottomrule
\endlastfoot
\addlinespace[2pt]
\textbf{instrument}$^{\dagger}$ & \texttt{object} & \textit{Details about the ITC instrument used.} \\
\addlinespace[1pt]
\textcolor{gray!70}{instrument.\allowbreak }\textbf{model}$^{\dagger}$ & \texttt{string} & Brand and model of the ITC instrument used. \emph{Examples: MicroCal VP-ITC; MicroCal PEAQ-ITC} \\
\addlinespace[2pt]
\textbf{protein}$^{\dagger}$ & \texttt{object} & \textit{Details about the protein/macromolecule used in the experiment.} \\
\addlinespace[1pt]
\textcolor{gray!70}{protein.\allowbreak }\textbf{name}$^{\dagger}$ & \texttt{string} & Name of the protein. \\
\textcolor{gray!70}{protein.\allowbreak }\textbf{location}$^{\dagger}$ & \texttt{string} & Physical location of the protein during the experiment. \emph{Allowed: cell; syringe} \\
\textcolor{gray!70}{protein.\allowbreak }\textbf{concentration}$^{\dagger}$ & \texttt{string?} & Concentration of the protein in micromolar (uM). Extract numeric value only (no units). Convert from other molar units: mM $\times$ 1000, nM / 1000. Use a string to allow ranges (e.g., '75-225'). Return null if the paper reports it in mass units (e.g., mg/mL) without providing a molecular weight for conversion. Use ASCII 'u' in unit strings, not Unicode micro. \emph{Examples: 15; 75-225; null} \\
\textcolor{gray!70}{protein.\allowbreak }\textbf{solution\_\allowbreak volume}$^{\dagger}$ & \texttt{string} & Volume of the protein solution in microliters (uL). Extract numeric value only. Convert from other units: mL $\times$ 1000. Use ASCII 'u' in unit strings, not Unicode micro. \emph{Examples: 3000; 270} \\
\addlinespace[2pt]
\textbf{ligand}$^{\dagger}$ & \texttt{object} & \textit{Details about the small molecule/ligand used in the experiment.} \\
\addlinespace[1pt]
\textcolor{gray!70}{ligand.\allowbreak }\textbf{name}$^{\dagger}$ & \texttt{string} & Name of the ligand. \\
\textcolor{gray!70}{ligand.\allowbreak }\textbf{location}$^{\dagger}$ & \texttt{string} & Physical location of the ligand during the experiment. \emph{Allowed: cell; syringe} \\
\textcolor{gray!70}{ligand.\allowbreak }\textbf{concentration}$^{\dagger}$ & \texttt{string?} & Concentration of the ligand in micromolar (uM). Extract numeric value only (no units). Convert from other molar units: mM $\times$ 1000, nM / 1000. Use a string to allow ranges (e.g., '10-30'). Return null if the paper reports it in mass units (e.g., mg/mL) without providing a molecular weight for conversion. Use ASCII 'u' in unit strings, not Unicode micro. \emph{Examples: 30; 10-30; null} \\
\addlinespace[2pt]
\textbf{assay\_\allowbreak conditions}$^{\dagger}$ & \texttt{object} & \textit{Conditions under which the titration was performed.} \\
\addlinespace[1pt]
\textcolor{gray!70}{assay\_\allowbreak conditions.\allowbreak }\textbf{buffer\_\allowbreak composition}$^{\dagger}$ & \texttt{string} & Composition of the buffer. Put concentration before chemical. Exclude pH. \emph{Examples: 10 mM Hepes, 150 mM NaCl, 3mM EDTA} \\
\textcolor{gray!70}{assay\_\allowbreak conditions.\allowbreak }\textbf{pH}$^{\dagger}$ & \texttt{number} & pH of the buffer. \emph{Examples: 7.4} \\
\textcolor{gray!70}{assay\_\allowbreak conditions.\allowbreak }\textbf{temperature\_\allowbreak c}$^{\dagger}$ & \texttt{string} & Temperature of the experiment in $^{\circ}$C. Extract numeric value only. \emph{Examples: 25} \\
\textcolor{gray!70}{assay\_\allowbreak conditions.\allowbreak }\textbf{stirring\_\allowbreak speed}$^{\dagger}$ & \texttt{string} & Stirring speed in rpm. Extract numeric value only. \emph{Examples: 1000} \\
\textcolor{gray!70}{assay\_\allowbreak conditions.\allowbreak }\textbf{reference\_\allowbreak cell\_\allowbreak content}$^{\dagger}$ & \texttt{string} & Content of the reference cell (usually water or buffer). \\
\textcolor{gray!70}{assay\_\allowbreak conditions.\allowbreak }\textbf{control\_\allowbreak method}$^{\dagger}$ & \texttt{string} & Description of control experiments/blanks performed to correct for heat of dilution. \emph{Examples: Titration of ligand into buffer; Titration of buffer into protein solution; Subtracted constant heat of injection} \\
\addlinespace[2pt]
\textcolor{gray!70}{assay\_\allowbreak conditions.\allowbreak }\textbf{injection\_\allowbreak parameters}$^{\dagger}$ & \texttt{object} & \textit{Specifics regarding the injection volumes and timing.} \\
\addlinespace[1pt]
\textcolor{gray!70}{assay\_\allowbreak conditions.\allowbreak injection\_\allowbreak parameters.\allowbreak }\textbf{first\_\allowbreak volume}$^{\dagger}$ & \texttt{string} & Volume of the first injection in microliters (uL). Extract numeric value only. Convert from other units: mL $\times$ 1000. Use ASCII 'u' in unit strings, not Unicode micro. \emph{Examples: 0.4; 1; 2} \\
\textcolor{gray!70}{assay\_\allowbreak conditions.\allowbreak injection\_\allowbreak parameters.\allowbreak }\textbf{subsequent\_\allowbreak volume}$^{\dagger}$ & \texttt{string} & Volume of subsequent injections in microliters (uL). Extract numeric value only. Convert from other units: mL $\times$ 1000. Use ASCII 'u' in unit strings, not Unicode micro. \emph{Examples: 2; 8; 10} \\
\textcolor{gray!70}{assay\_\allowbreak conditions.\allowbreak injection\_\allowbreak parameters.\allowbreak }\textbf{duration}$^{\dagger}$ & \texttt{string} & Duration of each injection in s. Extract numeric value only. \emph{Examples: 4; 16; 20} \\
\textcolor{gray!70}{assay\_\allowbreak conditions.\allowbreak injection\_\allowbreak parameters.\allowbreak }\textbf{spacing}$^{\dagger}$ & \texttt{string} & Time between each injection in s. Extract numeric value only. \emph{Examples: 120; 150; 180} \\
\textcolor{gray!70}{assay\_\allowbreak conditions.\allowbreak injection\_\allowbreak parameters.\allowbreak }\textbf{total\_\allowbreak count}$^{\dagger}$ & \texttt{string} & Total number of injections. Extract numeric value only. \emph{Examples: 19; 20; 25} \\
\addlinespace[2pt]
\textbf{data\_\allowbreak analysis}$^{\dagger}$ & \texttt{object} & \textit{Details about how the data was modeled.} \\
\addlinespace[1pt]
\textcolor{gray!70}{data\_\allowbreak analysis.\allowbreak }\textbf{binding\_\allowbreak model}$^{\dagger}$ & \texttt{string} & The binding model used to fit the ITC data. \emph{Examples: one-site; two-sites} \\
\end{longtable}
\end{small}

\subsection{Surface Plasmon Resonance (SPR)}
\label{app:schema:spr}
 
\begin{small}
\begin{longtable}{@{}>{\raggedright\arraybackslash}p{0.31\textwidth}>{\raggedright\arraybackslash}p{0.13\textwidth}>{\raggedright\arraybackslash}p{0.50\textwidth}@{}}
\caption{Schema ontology $\mathcal{O}$ for Surface Plasmon Resonance (SPR). A dagger ($\dagger$) marks fields required by the schema. \texttt{type?} denotes a nullable type.}\\
\label{tab:schema:spr} \\
\toprule
\textbf{Field path} & \textbf{Type} & \textbf{Description} \\
\midrule
\endfirsthead
\multicolumn{3}{c}{\textit{(continued from previous page)}} \\
\toprule
\textbf{Field path} & \textbf{Type} & \textbf{Description} \\
\midrule
\endhead
\midrule
\multicolumn{3}{r}{\textit{(continued on next page)}} \\
\endfoot
\bottomrule
\endlastfoot
\addlinespace[2pt]
\textbf{instrument}$^{\dagger}$ & \texttt{object} & \textit{Details about the SPR instrument used.} \\
\addlinespace[1pt]
\textcolor{gray!70}{instrument.\allowbreak }\textbf{manufacturer}$^{\dagger}$ & \texttt{string} & e.g., GE Healthcare, Cytiva, Biacore AB, PlexArrayTM \\
\textcolor{gray!70}{instrument.\allowbreak }\textbf{model}$^{\dagger}$ & \texttt{string} & e.g., Biacore T200, Biacore 8K, BIAcore 3000, PlexArrayTM HT SPRi \\
\addlinespace[2pt]
\textbf{sensor\_\allowbreak chip}$^{\dagger}$ & \texttt{object} & \textit{Details about the sensor chip.} \\
\addlinespace[1pt]
\textcolor{gray!70}{sensor\_\allowbreak chip.\allowbreak }\textbf{type}$^{\dagger}$ & \texttt{string} & The type of sensor chip surface. \emph{Examples: CM5; CM7; SA; NTA; ...} \\
\textcolor{gray!70}{sensor\_\allowbreak chip.\allowbreak }\textbf{manufacturer}$^{\dagger}$ & \texttt{string} & e.g., GE Healthcare, XanTec, Biacore \\
\addlinespace[2pt]
\textbf{immobilization}$^{\dagger}$ & \texttt{object} & \textit{Details about the molecule immobilized on the chip (the ligand).} \\
\addlinespace[1pt]
\textcolor{gray!70}{immobilization.\allowbreak }\textbf{ligand\_\allowbreak name}$^{\dagger}$ & \texttt{string} & The name of the protein, compound, or molecule attached to the chip. \emph{Examples: his-tagged HDAC8; Biotinylated MEK1; SARS-CoV-2 Mpro; Compound UA-18} \\
\textcolor{gray!70}{immobilization.\allowbreak }\textbf{strategy}$^{\dagger}$ & \texttt{string} & The chemical or affinity method used for immobilization. \emph{Allowed: Amine coupling; SA-Biotin capture; NTA-His capture; Thiol coupling; Aldehyde coupling; GST capture; Antibody capture; Photo-crosslinking; Direct immobilization; Other} \\
\textcolor{gray!70}{immobilization.\allowbreak }\textbf{density\_\allowbreak ru}$^{\dagger}$ & \texttt{string | number} & The immobilization level in Response Units (RU). Use a string to capture ranges or approximations. \emph{Examples: \textasciitilde{}4900; 6000-8000; 4000; 800-1800} \\
\textcolor{gray!70}{immobilization.\allowbreak }\textbf{concentration\_\allowbreak for\_\allowbreak immobilization}$^{\dagger}$ & \texttt{number?} & The concentration of the ligand used during the immobilization step, in micromolar (uM). Convert from other molar units: mM $\times$ 1000, nM / 1000. Return null if the paper reports it in mass units (e.g., ug/mL) without providing a molecular weight for conversion. Use ASCII 'u' in unit strings, not Unicode micro. \emph{Examples: 1; 50; 12.5; null} \\
\addlinespace[2pt]
\textbf{analyte}$^{\dagger}$ & \texttt{object} & \textit{Details about the molecule in solution being tested (the analyte).} \\
\addlinespace[1pt]
\textcolor{gray!70}{analyte.\allowbreak }\textbf{description}$^{\dagger}$ & \texttt{string} & A general description of the analyte(s). \emph{Examples: Fragments; Peptides; Small molecules; Compounds; ...} \\
\textcolor{gray!70}{analyte.\allowbreak }\textbf{concentration\_\allowbreak range}$^{\dagger}$ & \texttt{string} & The range of concentrations tested, described as text. Normalize all concentrations to micromolar (uM) using: mM $\times$ 1000, nM / 1000. Use ASCII 'u' in unit strings, not Unicode micro. \emph{Examples: 2000 uM; 5-0.0195 uM; 0.2-100 uM; 9-step, 2-fold, top 0.02 uM} \\
\addlinespace[2pt]
\textbf{assay\_\allowbreak conditions}$^{\dagger}$ & \texttt{object} & \textit{Conditions during the binding (association/dissociation) phase.} \\
\addlinespace[1pt]
\textcolor{gray!70}{assay\_\allowbreak conditions.\allowbreak }\textbf{running\_\allowbreak buffer\_\allowbreak composition}$^{\dagger}$ & \texttt{string} & The full composition of the running buffer, including additives (e.g., DMSO, Tween). Exclude pH here; use the separate pH field. \\
\textcolor{gray!70}{assay\_\allowbreak conditions.\allowbreak }\textbf{pH}$^{\dagger}$ & \texttt{number} & The pH of the running buffer. \emph{Examples: 7.4; 6.0; 7.5} \\
\textcolor{gray!70}{assay\_\allowbreak conditions.\allowbreak }\textbf{assay\_\allowbreak type}$^{\dagger}$ & \texttt{string} & The type of kinetic experiment performed. \emph{Allowed: Single Cycle Kinetics; Multi Cycle Kinetics; Single Cycle Steady-State Affinity; Multi Cycle Steady-State Affinity; Screening; Unknown} \\
\textcolor{gray!70}{assay\_\allowbreak conditions.\allowbreak }\textbf{assay\_\allowbreak flow\_\allowbreak rate}$^{\dagger}$ & \texttt{number} & The flow rate used during analyte injection, in microliters per minute (uL/min). \\
\textcolor{gray!70}{assay\_\allowbreak conditions.\allowbreak }\textbf{temperature\_\allowbreak c}$^{\dagger}$ & \texttt{string} & The temperature of the experiment. Use a string to allow for 'room temperature'. \emph{Examples: 25; 4; room temperature} \\
\textcolor{gray!70}{assay\_\allowbreak conditions.\allowbreak }\textbf{association\_\allowbreak time\_\allowbreak s}$^{\dagger}$ & \texttt{string | number} & The contact/association time in seconds. Use a string to capture ranges. \emph{Examples: 60; 120; 70; 30-600} \\
\textcolor{gray!70}{assay\_\allowbreak conditions.\allowbreak }\textbf{dissociation\_\allowbreak time\_\allowbreak s}$^{\dagger}$ & \texttt{string | number} & The dissociation time in seconds. Use a string to capture ranges. \emph{Examples: 240; 420; 600; 100-600; ...} \\
\textcolor{gray!70}{assay\_\allowbreak conditions.\allowbreak }\textbf{regeneration\_\allowbreak solution}$^{\dagger}$ & \texttt{string} & The composition of the solution used to regenerate the chip surface. \emph{Examples: None; 10 mM glycine pH 2; 40 mM CHAPS...; 9:1 1M NaCl: Glycerol} \\
\addlinespace[2pt]
\textbf{data\_\allowbreak analysis}$^{\dagger}$ & \texttt{object} & \textit{Details about how the raw data was processed and modeled.} \\
\addlinespace[1pt]
\textcolor{gray!70}{data\_\allowbreak analysis.\allowbreak }\textbf{reference\_\allowbreak description}$^{\dagger}$ & \texttt{string} & Description of the reference surface or channel used for subtraction. \emph{Examples: Untreated SA surface; Fc1 (no ligand); Reference channels; Flow-cells 1 and 3} \\
\textcolor{gray!70}{data\_\allowbreak analysis.\allowbreak }\textbf{subtraction\_\allowbreak method}$^{\dagger}$ & \texttt{string} & The method of data normalization. \emph{Allowed: Double referenced; Reference subtracted; Solvent correction; Background subtracted; Blank subtracted; Other} \\
\textcolor{gray!70}{data\_\allowbreak analysis.\allowbreak }\textbf{fitting\_\allowbreak model}$^{\dagger}$ & \texttt{string} & The kinetic or affinity model used to fit the data. \emph{Allowed: 1:1 Langmuir binding model; 1:1 binding model with mass transport; Steady-state affinity; Two-state binding model; Heterogeneous binding model; Bivalent analyte model; Other} \\
\textcolor{gray!70}{data\_\allowbreak analysis.\allowbreak }\textbf{software}$^{\dagger}$ & \texttt{string} & The software used for data analysis. \emph{Examples: BiaEvaluation 3.0; GeneData; Scrubber software, version 2.0c; Biacore 8K Insight Evaluation Software Version 2.0} \\
\end{longtable}
\end{small}
 
\subsection{Radioligand Binding Assay (RBA)}
\label{app:schema:rba}
 
\begin{small}
\begin{longtable}{@{}>{\raggedright\arraybackslash}p{0.31\textwidth}>{\raggedright\arraybackslash}p{0.13\textwidth}>{\raggedright\arraybackslash}p{0.50\textwidth}@{}}
\caption{Schema ontology $\mathcal{O}$ for Radioligand Binding Assay (RBA). A dagger ($\dagger$) marks fields required by the schema. \texttt{type?} denotes a nullable type.}\\
\label{tab:schema:rba} \\
\toprule
\textbf{Field path} & \textbf{Type} & \textbf{Description} \\
\midrule
\endfirsthead
\multicolumn{3}{c}{\textit{(continued from previous page)}} \\
\toprule
\textbf{Field path} & \textbf{Type} & \textbf{Description} \\
\midrule
\endhead
\midrule
\multicolumn{3}{r}{\textit{(continued on next page)}} \\
\endfoot
\bottomrule
\endlastfoot
\addlinespace[2pt]
\textbf{biological\_\allowbreak preparation}$^{\dagger}$ & \texttt{object} & \textit{Details about the biological source containing the target receptor.} \\
\addlinespace[1pt]
\textcolor{gray!70}{biological\_\allowbreak preparation.\allowbreak }\textbf{source\_\allowbreak type}$^{\dagger}$ & \texttt{string} & The type of biological material used. \emph{Allowed: Membrane homogenate; Whole cells; Tissue section; Purified protein; Cell lysate; Tissue homogenate; Other} \\
\textcolor{gray!70}{biological\_\allowbreak preparation.\allowbreak }\textbf{tissue\_\allowbreak or\_\allowbreak cell\_\allowbreak line}$^{\dagger}$ & \texttt{string} & The tissue, cell line, or expression system used. \emph{Examples: Rat striatal membranes; CHO-K1 cells stably expressing hD2R; HEK293 cells transfected with hMOR; Guinea pig brain cortex; ...} \\
\textcolor{gray!70}{biological\_\allowbreak preparation.\allowbreak }\textbf{species}$^{\dagger}$ & \texttt{string} & The species origin of the biological preparation. \emph{Examples: Human; Rat; Mouse; Guinea pig; ...} \\
\textcolor{gray!70}{biological\_\allowbreak preparation.\allowbreak }\textbf{target\_\allowbreak protein}$^{\dagger}$ & \texttt{string} & The receptor or protein target being studied, including isoform or subunit composition if specified. \emph{Examples: Dopamine D2 receptor; GABAA ($\alpha$1$\beta$2$\gamma$2); Muscarinic M1 receptor; Opioid mu receptor (MOR); ...} \\
\textcolor{gray!70}{biological\_\allowbreak preparation.\allowbreak }\textbf{protein\_\allowbreak concentration}$^{\dagger}$ & \texttt{string} & The concentration of membrane protein or cells per well/tube, including units. Heterogeneous field: may be mass-based (ug/well, ug/tube, ug/mL) or cell-count-based (cells/well). Use ASCII 'u' in unit strings, not Unicode micro. \emph{Examples: 50 ug/well; 20 ug protein/tube; 100,000 cells/well; 15 ug membrane protein} \\
\addlinespace[2pt]
\textbf{radioligand}$^{\dagger}$ & \texttt{object} & \textit{Details about the radiolabeled probe used in the assay.} \\
\addlinespace[1pt]
\textcolor{gray!70}{radioligand.\allowbreak }\textbf{name}$^{\dagger}$ & \texttt{string} & The name of the radioligand, including the isotope label. \emph{Examples: [3H]-SCH 23390; [125I]-RTI-55; [3H]-DPCPX; [3H]-raclopride; ...} \\
\textcolor{gray!70}{radioligand.\allowbreak }\textbf{isotope}$^{\dagger}$ & \texttt{string} & The radioisotope used for labeling. \emph{Allowed: 3H; 125I; 35S; 14C; 33P; Other} \\
\addlinespace[2pt]
\textcolor{gray!70}{radioligand.\allowbreak }\textbf{specific\_\allowbreak activity}$^{\dagger}$ & \texttt{object} & \textit{The specific activity of the radioligand.} \\
\addlinespace[1pt]
\textcolor{gray!70}{radioligand.\allowbreak specific\_\allowbreak activity.\allowbreak }\textbf{value}$^{\dagger}$ & \texttt{string | number} & Numeric value of the specific activity. Use a string to capture ranges or approximations. \emph{Examples: 85.5; 81-86; 2200} \\
\textcolor{gray!70}{radioligand.\allowbreak specific\_\allowbreak activity.\allowbreak }\textbf{unit}$^{\dagger}$ & \texttt{string} & Unit of specific activity. \emph{Allowed: Ci/mmol; mCi/mmol; GBq/mmol; TBq/mmol} \\
\textcolor{gray!70}{radioligand.\allowbreak }\textbf{concentration\_\allowbreak used}$^{\dagger}$ & \texttt{number?} & The concentration of radioligand used in the assay, as a number in nanomolar (nM). Convert from other molar units: pM / 1000, uM $\times$ 1000, mM $\times$ 1000000. Return null if the paper reports only a qualitative value (e.g., '\textasciitilde{}Kd concentration') or uses units that cannot be converted to nM. Use ASCII 'u' in unit strings elsewhere in the schema, not Unicode micro. \emph{Examples: 1; 0.5; 2; null} \\
\addlinespace[2pt]
\textcolor{gray!70}{radioligand.\allowbreak }\textbf{kd\_\allowbreak value}$^{\dagger}$ & \texttt{object} & \textit{The known dissociation constant (Kd) of the radioligand for the target, required for Ki calculation via Cheng-Prusoff.} \\
\addlinespace[1pt]
\textcolor{gray!70}{radioligand.\allowbreak kd\_\allowbreak value.\allowbreak }\textbf{value}$^{\dagger}$ & \texttt{number} & Kd value as a number in nanomolar (nM). Convert from other units: uM $\times$ 1000, pM / 1000. Use ASCII 'u' in unit strings, not Unicode micro. \\
\textcolor{gray!70}{radioligand.\allowbreak kd\_\allowbreak value.\allowbreak }\textbf{source}$^{\dagger}$ & \texttt{string} & Whether the Kd was determined in-house or taken from literature. \emph{Allowed: Determined in this study; Literature value; Not specified} \\
\addlinespace[2pt]
\textbf{test\_\allowbreak compound}$^{\dagger}$ & \texttt{object} & \textit{Details about the unlabeled compound(s) being evaluated.} \\
\addlinespace[1pt]
\textcolor{gray!70}{test\_\allowbreak compound.\allowbreak }\textbf{description}$^{\dagger}$ & \texttt{string} & A general description of the test compound(s). \emph{Examples: Small molecules; Peptides; Novel D2 antagonists; Natural product library} \\
\textcolor{gray!70}{test\_\allowbreak compound.\allowbreak }\textbf{concentration\_\allowbreak range}$^{\dagger}$ & \texttt{string} & The range of concentrations tested for the competition curve, as a string with all concentrations normalized to micromolar (uM). Convert from other molar units: nM / 1000, pM / 1000000, mM $\times$ 1000. Use ASCII 'u' in unit strings, not Unicode micro. \emph{Examples: 0.0001-10 uM; 10-point, half-log dilution, top 10 uM; 12 concentrations, 3-fold serial dilution; Single point at 10 uM} \\
\textcolor{gray!70}{test\_\allowbreak compound.\allowbreak }\textbf{vehicle\_\allowbreak solvent}$^{\dagger}$ & \texttt{string} & The solvent used to dissolve the test compounds and its final assay concentration. \emph{Examples: DMSO, final 1\%; DMSO, final 0.1\%; Water; Ethanol, final 0.5\%} \\
\addlinespace[2pt]
\textbf{nsb\_\allowbreak definition}$^{\dagger}$ & \texttt{object} & \textit{How nonspecific binding (NSB) was defined in the assay. This is critical for data interpretation.} \\
\addlinespace[1pt]
\textcolor{gray!70}{nsb\_\allowbreak definition.\allowbreak }\textbf{compound\_\allowbreak name}$^{\dagger}$ & \texttt{string} & The unlabeled compound used to define 100\% displacement (nonspecific binding). \emph{Examples: SCH 23390; Haloperidol; Naloxone; Atropine; ...} \\
\textcolor{gray!70}{nsb\_\allowbreak definition.\allowbreak }\textbf{compound\_\allowbreak concentration}$^{\dagger}$ & \texttt{number?} & The concentration of the NSB-defining compound as a number in micromolar (uM). Convert from other molar units: nM / 1000, mM $\times$ 1000. Return null if the paper uses units that cannot be converted to uM. Use ASCII 'u' in unit strings, not Unicode micro. \emph{Examples: 10; 1; 100; null} \\
\textcolor{gray!70}{nsb\_\allowbreak definition.\allowbreak }\textbf{selectivity\_\allowbreak rationale}$^{\dagger}$ & \texttt{string} & Whether the NSB compound is selective for the target, or a broad/non-selective displacer. \emph{Allowed: Target-selective; Non-selective; Not specified} \\
\addlinespace[2pt]
\textbf{assay\_\allowbreak conditions}$^{\dagger}$ & \texttt{object} & \textit{Conditions during the incubation phase.} \\
\addlinespace[1pt]
\textcolor{gray!70}{assay\_\allowbreak conditions.\allowbreak }\textbf{assay\_\allowbreak format}$^{\dagger}$ & \texttt{string} & The physical format of the assay. \emph{Examples: Filtration (well); Filtration (tubes); Scintillation Proximity Assay (SPA); Centrifugation; ...} \\
\textcolor{gray!70}{assay\_\allowbreak conditions.\allowbreak }\textbf{buffer\_\allowbreak composition}$^{\dagger}$ & \texttt{string} & The full composition of the incubation buffer, including salts and additives. Exclude pH here; use the separate pH field. \emph{Examples: 50 mM Tris-HCl, 120 mM NaCl, 5 mM MgCl2, 1 mM EDTA; 50 mM HEPES, 10 mM MgCl2, 100 mM NaCl} \\
\textcolor{gray!70}{assay\_\allowbreak conditions.\allowbreak }\textbf{pH}$^{\dagger}$ & \texttt{number} & The pH of the incubation buffer. \emph{Examples: 7.4; 7.5; 6.8} \\
\textcolor{gray!70}{assay\_\allowbreak conditions.\allowbreak }\textbf{incubation\_\allowbreak temperature}$^{\dagger}$ & \texttt{string} & The temperature during incubation. Use a string to allow for 'room temperature'. \emph{Examples: 25; 37; 4; room temperature} \\
\textcolor{gray!70}{assay\_\allowbreak conditions.\allowbreak }\textbf{incubation\_\allowbreak time}$^{\dagger}$ & \texttt{string | number} & The duration of incubation as a number in minutes. Convert from other units: h $\times$ 60, s / 60. Use a string to capture ranges (e.g., '60-120'). \emph{Examples: 60; 90; 120; 60-120} \\
\textcolor{gray!70}{assay\_\allowbreak conditions.\allowbreak }\textbf{equilibrium\_\allowbreak confirmed}$^{\dagger}$ & \texttt{string} & Whether the authors confirmed that binding had reached equilibrium (e.g., via a time course). \emph{Allowed: Yes; No; Not specified} \\
\textcolor{gray!70}{assay\_\allowbreak conditions.\allowbreak }\textbf{total\_\allowbreak assay\_\allowbreak volume}$^{\dagger}$ & \texttt{number} & The total volume of the assay mixture as a number in microliters (uL). Convert from other units: mL $\times$ 1000. Use ASCII 'u' in unit strings, not Unicode micro. \emph{Examples: 200; 500; 1000; 250} \\
\addlinespace[2pt]
\textbf{separation\_\allowbreak and\_\allowbreak detection}$^{\dagger}$ & \texttt{object} & \textit{How bound and free radioligand were separated and how radioactivity was measured.} \\
\addlinespace[1pt]
\textcolor{gray!70}{separation\_\allowbreak and\_\allowbreak detection.\allowbreak }\textbf{separation\_\allowbreak method}$^{\dagger}$ & \texttt{string} & The method used to separate bound from free radioligand. \emph{Allowed: Rapid vacuum filtration; Harvester filtration; Centrifugation; SPA (no separation); FlashPlate (no separation); Other} \\
\textcolor{gray!70}{separation\_\allowbreak and\_\allowbreak detection.\allowbreak }\textbf{filter\_\allowbreak type}$^{\dagger}$ & \texttt{string?} & The type of filter used, if filtration-based. Null if not applicable. \emph{Examples: GF/B; GF/C; Unifilter-96 GF/B; Whatman GF/B pretreated with 0.3\% PEI} \\
\textcolor{gray!70}{separation\_\allowbreak and\_\allowbreak detection.\allowbreak }\textbf{wash\_\allowbreak protocol}$^{\dagger}$ & \texttt{string?} & The wash conditions after filtration. Null if not applicable. \emph{Examples: 3 $\times$ 3 mL ice-cold 50 mM Tris-HCl pH 7.4; 4 $\times$ 1 mL cold wash buffer; 3 $\times$ 500 uL ice-cold buffer} \\
\textcolor{gray!70}{separation\_\allowbreak and\_\allowbreak detection.\allowbreak }\textbf{detection\_\allowbreak instrument}$^{\dagger}$ & \texttt{string} & The instrument used to measure radioactivity. \emph{Examples: Perkin Elmer MicroBeta2; Beckman LS6500 scintillation counter; Packard TopCount NXT; Wallac 1450 MicroBeta} \\
\textcolor{gray!70}{separation\_\allowbreak and\_\allowbreak detection.\allowbreak }\textbf{scintillation\_\allowbreak cocktail}$^{\dagger}$ & \texttt{string?} & The scintillation cocktail used for liquid scintillation counting. Null if gamma counting or proximity-based. \emph{Examples: Ultima Gold; MicroScint-20; EcoLite; OptiPhase SuperMix} \\
\addlinespace[2pt]
\textbf{data\_\allowbreak analysis}$^{\dagger}$ & \texttt{object} & \textit{Details about how the raw data was processed and modeled.} \\
\addlinespace[1pt]
\textcolor{gray!70}{data\_\allowbreak analysis.\allowbreak }\textbf{endpoint\_\allowbreak reported}$^{\dagger}$ & \texttt{string} & The primary binding parameter reported in the publication. \emph{Allowed: Ki; IC50; Kd; Bmax; Percent inhibition; pKi; pIC50; EC50; Other} \\
\textcolor{gray!70}{data\_\allowbreak analysis.\allowbreak }\textbf{ic50\_\allowbreak to\_\allowbreak ki\_\allowbreak conversion}$^{\dagger}$ & \texttt{string?} & The method used to convert IC50 to Ki, if applicable. Null if Ki was not derived from IC50. \emph{Allowed: Cheng-Prusoff equation; Munson-Rodbard correction; Exact solution (Cer \& Bhatt); Not applicable; Not specified} \\
\textcolor{gray!70}{data\_\allowbreak analysis.\allowbreak }\textbf{fitting\_\allowbreak model}$^{\dagger}$ & \texttt{string} & The model used to fit competition or saturation binding curves. \emph{Examples: One-site competition; Two-site competition; Four-parameter logistic; Saturation binding (one-site); ...} \\
\addlinespace[2pt]
\textcolor{gray!70}{data\_\allowbreak analysis.\allowbreak }\textbf{hill\_\allowbreak slope\_\allowbreak reported}$^{\dagger}$ & \texttt{object} & \textit{Whether a Hill slope (nH) was reported and its value, if available. nH $\neq$ 1 may indicate cooperativity or binding heterogeneity.} \\
\addlinespace[1pt]
\textcolor{gray!70}{data\_\allowbreak analysis.\allowbreak hill\_\allowbreak slope\_\allowbreak reported.\allowbreak }\textbf{reported}$^{\dagger}$ & \texttt{string} &  \emph{Allowed: Yes, fixed to 1; Yes, variable; Not reported} \\
\textcolor{gray!70}{data\_\allowbreak analysis.\allowbreak hill\_\allowbreak slope\_\allowbreak reported.\allowbreak }\textbf{value} & \texttt{number?} & The Hill slope value if reported as variable. Null if fixed or not reported. \\
\textcolor{gray!70}{data\_\allowbreak analysis.\allowbreak }\textbf{software}$^{\dagger}$ & \texttt{string} & The software used for curve fitting and data analysis. \emph{Examples: GraphPad Prism 9; GraphPad Prism 8; SigmaPlot; GeneData Screener; ...} \\
\addlinespace[2pt]
\textcolor{gray!70}{data\_\allowbreak analysis.\allowbreak }\textbf{replicates}$^{\dagger}$ & \texttt{object} & \textit{The number of replicates and independent experiments.} \\
\addlinespace[1pt]
\textcolor{gray!70}{data\_\allowbreak analysis.\allowbreak replicates.\allowbreak }\textbf{technical\_\allowbreak replicates}$^{\dagger}$ & \texttt{string} & Number of replicates per concentration within a single experiment (e.g., duplicate, triplicate). \emph{Examples: Duplicate; Triplicate; Singlicate; Not specified} \\
\textcolor{gray!70}{data\_\allowbreak analysis.\allowbreak replicates.\allowbreak }\textbf{independent\_\allowbreak experiments}$^{\dagger}$ & \texttt{string | number} & Number of independent experiments (n). Use a string to capture ranges. \emph{Examples: 3; 2-4; $\geq$3; Not specified} \\
\end{longtable}
\end{small}
 
\subsection{Fluorescence Polarization / Anisotropy (FPA)}
\label{app:schema:fpa}
 
\begin{small}
\begin{longtable}{@{}>{\raggedright\arraybackslash}p{0.31\textwidth}>{\raggedright\arraybackslash}p{0.13\textwidth}>{\raggedright\arraybackslash}p{0.50\textwidth}@{}}
\caption{Schema ontology $\mathcal{O}$ for Fluorescence Polarization / Anisotropy (FPA). A dagger ($\dagger$) marks fields required by the schema. \texttt{type?} denotes a nullable type.}\\
\label{tab:schema:fpa} \\
\toprule
\textbf{Field path} & \textbf{Type} & \textbf{Description} \\
\midrule
\endfirsthead
\multicolumn{3}{c}{\textit{(continued from previous page)}} \\
\toprule
\textbf{Field path} & \textbf{Type} & \textbf{Description} \\
\midrule
\endhead
\midrule
\multicolumn{3}{r}{\textit{(continued on next page)}} \\
\endfoot
\bottomrule
\endlastfoot
\addlinespace[2pt]
\textbf{biological\_\allowbreak preparation}$^{\dagger}$ & \texttt{object} & \textit{Details about the biological source containing the target protein.} \\
\addlinespace[1pt]
\textcolor{gray!70}{biological\_\allowbreak preparation.\allowbreak }\textbf{source\_\allowbreak type}$^{\dagger}$ & \texttt{string} & The type of biological material used. \emph{Examples: Purified protein; Membrane homogenate; Whole cells; Cell lysate; ...} \\
\textcolor{gray!70}{biological\_\allowbreak preparation.\allowbreak }\textbf{tissue\_\allowbreak or\_\allowbreak cell\_\allowbreak line}$^{\dagger}$ & \texttt{string} & The tissue, cell line, or expression system used to produce the protein. \emph{Examples: E. coli BL21(DE3) expressing His-tagged hBRD4-BD1; CHO-K1 cells stably expressing hER$\alpha$-LBD; Sf9 insect cells expressing hHsp90$\alpha$; HEK293T cells transfected with hMDM2} \\
\textcolor{gray!70}{biological\_\allowbreak preparation.\allowbreak }\textbf{species}$^{\dagger}$ & \texttt{string} & The species origin of the target protein. \emph{Examples: Human; Rat; Mouse; Bovine; ...} \\
\textcolor{gray!70}{biological\_\allowbreak preparation.\allowbreak }\textbf{target\_\allowbreak protein}$^{\dagger}$ & \texttt{string} & The protein target being studied, including domain or construct boundaries if specified. \emph{Examples: BRD4 bromodomain 1 (BD1, residues 44--168); MDM2 (residues 1--118); Hsp90$\alpha$ N-terminal domain; ER$\alpha$ ligand-binding domain; ...} \\
\textcolor{gray!70}{biological\_\allowbreak preparation.\allowbreak }\textbf{protein\_\allowbreak concentration}$^{\dagger}$ & \texttt{number | string?} & The concentration of protein used in the assay, as a number in nanomolar (nM). Convert from other molar units: uM * 1000, pM / 1000, mM * 1000000. Return null if the paper reports mass units (e.g., mg/mL) without a molecular weight for conversion, or if not stated. Use ASCII 'u', not Unicode micro. \emph{Examples: 50; 100; 25; 200; ...} \\
\textcolor{gray!70}{biological\_\allowbreak preparation.\allowbreak }\textbf{protein\_\allowbreak purity\_\allowbreak method} & \texttt{string?} & The purification method used, if reported. \emph{Examples: Ni-NTA affinity chromatography followed by SEC; GST pulldown and TEV cleavage; Not specified} \\
\addlinespace[2pt]
\textbf{fluorescent\_\allowbreak tracer}$^{\dagger}$ & \texttt{object} & \textit{Details about the fluorescently labeled probe used in the assay. Analogous to the radioligand in an RBA.} \\
\addlinespace[1pt]
\textcolor{gray!70}{fluorescent\_\allowbreak tracer.\allowbreak }\textbf{name}$^{\dagger}$ & \texttt{string} & The name or identifier of the fluorescent tracer, including the fluorophore label. \emph{Examples: FITC-labeled p53 peptide (residues 15--29); TAMRA-JQ1; FAM-SMAC peptide (AVPIAQK); BODIPY-geldanamycin; ...} \\
\textcolor{gray!70}{fluorescent\_\allowbreak tracer.\allowbreak }\textbf{tracer\_\allowbreak type}$^{\dagger}$ & \texttt{string} & The chemical nature of the tracer. \emph{Examples: Labeled peptide; Labeled small molecule; Labeled nucleic acid; Labeled protein fragment; ...} \\
\textcolor{gray!70}{fluorescent\_\allowbreak tracer.\allowbreak }\textbf{fluorophore}$^{\dagger}$ & \texttt{string} & The fluorophore conjugated to the tracer. \emph{Examples: Fluorescein (FITC); FAM (6-carboxyfluorescein); TAMRA; BODIPY; ...} \\
\textcolor{gray!70}{fluorescent\_\allowbreak tracer.\allowbreak }\textbf{excitation\_\allowbreak wavelength\_\allowbreak nm} & \texttt{number?} & The excitation wavelength used, in nanometers. \emph{Examples: 485; 530; 635; 480} \\
\textcolor{gray!70}{fluorescent\_\allowbreak tracer.\allowbreak }\textbf{emission\_\allowbreak wavelength\_\allowbreak nm} & \texttt{number?} & The emission wavelength used, in nanometers. \emph{Examples: 530; 580; 670; 520} \\
\textcolor{gray!70}{fluorescent\_\allowbreak tracer.\allowbreak }\textbf{concentration\_\allowbreak used}$^{\dagger}$ & \texttt{number?} & The concentration of tracer used in the assay, as a number in nanomolar (nM). Convert from other molar units: uM * 1000, pM / 1000. Return null if only qualitative (e.g., '\textasciitilde{}Kd concentration') or if not stated. Use ASCII 'u', not Unicode micro. \emph{Examples: 5; 10; 1; 20} \\
\addlinespace[2pt]
\textcolor{gray!70}{fluorescent\_\allowbreak tracer.\allowbreak }\textbf{kd\_\allowbreak value}$^{\dagger}$ & \texttt{object} & \textit{The known dissociation constant (Kd) of the tracer for the target protein, required for Ki calculation via Cheng-Prusoff.} \\
\addlinespace[1pt]
\textcolor{gray!70}{fluorescent\_\allowbreak tracer.\allowbreak kd\_\allowbreak value.\allowbreak }\textbf{value}$^{\dagger}$ & \texttt{number?} & Kd as a number in nanomolar (nM). Convert from other molar units: uM * 1000, pM / 1000. Use ASCII 'u', not Unicode micro. \\
\textcolor{gray!70}{fluorescent\_\allowbreak tracer.\allowbreak kd\_\allowbreak value.\allowbreak }\textbf{source}$^{\dagger}$ & \texttt{string} & Whether the tracer Kd was determined in-house or taken from literature. \emph{Examples: Determined in this study; Literature value; Not specified} \\
\addlinespace[2pt]
\textbf{test\_\allowbreak compound}$^{\dagger}$ & \texttt{object} & \textit{Details about the unlabeled competitor compound(s) being evaluated.} \\
\addlinespace[1pt]
\textcolor{gray!70}{test\_\allowbreak compound.\allowbreak }\textbf{description}$^{\dagger}$ & \texttt{string} & A general description of the test compound(s). \emph{Examples: Small molecules; Peptides; Novel BRD4 inhibitors; Stapled peptides targeting MDM2; ...} \\
\textcolor{gray!70}{test\_\allowbreak compound.\allowbreak }\textbf{concentration\_\allowbreak range}$^{\dagger}$ & \texttt{string} & The range of concentrations tested for the competition curve, as a string with all concentrations normalized to micromolar (uM). Convert from other molar units: mM * 1000, nM / 1000, pM / 1000000. Use ASCII 'u', not Unicode micro. \emph{Examples: 0.0001-100 uM; 10-point, half-log dilution, top 50 uM; 12 concentrations, 3-fold serial dilution; Single point at 10 uM} \\
\textcolor{gray!70}{test\_\allowbreak compound.\allowbreak }\textbf{vehicle\_\allowbreak solvent}$^{\dagger}$ & \texttt{string} & The solvent used to dissolve the test compounds and its final assay concentration. \emph{Examples: DMSO, final 1\%; DMSO, final 0.5\%; DMSO, final 2\%; Water} \\
\textcolor{gray!70}{test\_\allowbreak compound.\allowbreak }\textbf{dmso\_\allowbreak tolerance\_\allowbreak tested} & \texttt{string} & Whether the authors verified that the final DMSO concentration does not interfere with FP signal or protein-tracer binding. \emph{Allowed: Yes; No; Not specified} \\
\addlinespace[2pt]
\textbf{assay\_\allowbreak controls}$^{\dagger}$ & \texttt{object} & \textit{How the assay window (fully bound and fully displaced signals) was defined. Analogous to NSB definition in RBA.} \\
\addlinespace[1pt]
\textcolor{gray!70}{assay\_\allowbreak controls.\allowbreak }\textbf{high\_\allowbreak signal\_\allowbreak control}$^{\dagger}$ & \texttt{string} & The condition representing maximum polarization/anisotropy (fully bound tracer). \emph{Examples: Protein + tracer, no competitor; Protein + tracer + DMSO vehicle} \\
\textcolor{gray!70}{assay\_\allowbreak controls.\allowbreak }\textbf{low\_\allowbreak signal\_\allowbreak control}$^{\dagger}$ & \texttt{string} & The condition representing minimum polarization/anisotropy (fully displaced or free tracer). \emph{Examples: Tracer only, no protein; Tracer + 100 uM unlabeled JQ1; Tracer + 50 uM reference competitor} \\
\textcolor{gray!70}{assay\_\allowbreak controls.\allowbreak }\textbf{positive\_\allowbreak control\_\allowbreak compound} & \texttt{string?} & A reference compound of known affinity used to validate the assay. \emph{Examples: (+)-JQ1; Nutlin-3a; 17-AAG; GDC-0152; ...} \\
\textcolor{gray!70}{assay\_\allowbreak controls.\allowbreak }\textbf{positive\_\allowbreak control\_\allowbreak expected\_\allowbreak value} & \texttt{string?} & The expected Ki or IC50 for the positive control compound, including units. \emph{Examples: Ki = 50 nM; IC50 = 0.6 uM; Not specified} \\
\addlinespace[2pt]
\textcolor{gray!70}{assay\_\allowbreak controls.\allowbreak }\textbf{z\_\allowbreak prime\_\allowbreak reported}$^{\dagger}$ & \texttt{object} & \textit{Whether the Z'-factor (assay quality metric) was reported and its value.} \\
\addlinespace[1pt]
\textcolor{gray!70}{assay\_\allowbreak controls.\allowbreak z\_\allowbreak prime\_\allowbreak reported.\allowbreak }\textbf{reported}$^{\dagger}$ & \texttt{string} &  \emph{Allowed: Yes; No} \\
\textcolor{gray!70}{assay\_\allowbreak controls.\allowbreak z\_\allowbreak prime\_\allowbreak reported.\allowbreak }\textbf{value} & \texttt{number?} & The Z'-factor value, if reported. Null otherwise. \\
\addlinespace[2pt]
\textbf{assay\_\allowbreak conditions}$^{\dagger}$ & \texttt{object} & \textit{Conditions during the incubation phase.} \\
\addlinespace[1pt]
\textcolor{gray!70}{assay\_\allowbreak conditions.\allowbreak }\textbf{assay\_\allowbreak format}$^{\dagger}$ & \texttt{string} & The plate format used. \emph{Allowed: 384-well black plate; 96-well black plate; 384-well low-volume black plate; 1536-well black plate; Other} \\
\textcolor{gray!70}{assay\_\allowbreak conditions.\allowbreak }\textbf{plate\_\allowbreak type} & \texttt{string?} & Specific plate information if reported. \emph{Examples: Corning 3575 (384-well, low-volume, black, flat-bottom); Greiner 784076 (384-well, black, flat-bottom); PerkinElmer OptiPlate-384; Not specified} \\
\textcolor{gray!70}{assay\_\allowbreak conditions.\allowbreak }\textbf{buffer\_\allowbreak composition}$^{\dagger}$ & \texttt{string} & The full composition of the assay buffer, including salts and additives. Exclude pH here; use the separate pH field. \emph{Examples: 50 mM HEPES, 150 mM NaCl, 0.01\% Triton X-100, 1 mM DTT; 100 mM phosphate buffer, 100 mM NaCl, 0.05\% CHAPS; PBS with 0.01\% Tween-20 and 0.1 mg/mL BSA} \\
\textcolor{gray!70}{assay\_\allowbreak conditions.\allowbreak }\textbf{ph}$^{\dagger}$ & \texttt{number} & The pH of the assay buffer. \emph{Examples: 7.4; 7.5; 7.0} \\
\textcolor{gray!70}{assay\_\allowbreak conditions.\allowbreak }\textbf{detergent\_\allowbreak used} & \texttt{string?} & Whether a detergent was included to reduce compound aggregation or nonspecific binding, and which one. \emph{Examples: 0.01\% Triton X-100; 0.05\% CHAPS; 0.01\% Tween-20; 0.05\% Pluronic F-127; ...} \\
\textcolor{gray!70}{assay\_\allowbreak conditions.\allowbreak }\textbf{incubation\_\allowbreak temperature}$^{\dagger}$ & \texttt{string} & The temperature during incubation. Use a string to allow for 'room temperature'. \emph{Examples: 25; 37; 4; room temperature} \\
\textcolor{gray!70}{assay\_\allowbreak conditions.\allowbreak }\textbf{incubation\_\allowbreak time}$^{\dagger}$ & \texttt{string | number} & The duration of incubation before reading, as a number in minutes. Convert from other units: h * 60. Use a string for ranges (e.g., '30-60') or qualitative values (e.g., 'overnight'). \emph{Examples: 30; 60; 120; 30-60; ...} \\
\textcolor{gray!70}{assay\_\allowbreak conditions.\allowbreak }\textbf{equilibrium\_\allowbreak confirmed}$^{\dagger}$ & \texttt{string} & Whether the authors confirmed that binding had reached equilibrium (e.g., via a time course of FP signal). \emph{Allowed: Yes; No; Not specified} \\
\textcolor{gray!70}{assay\_\allowbreak conditions.\allowbreak }\textbf{total\_\allowbreak assay\_\allowbreak volume}$^{\dagger}$ & \texttt{number?} & The total volume of the assay mixture per well, as a number in microliters (uL). Convert from other units: mL * 1000. Use ASCII 'u', not Unicode micro. \emph{Examples: 20; 25; 50; 100; ...} \\
\textcolor{gray!70}{assay\_\allowbreak conditions.\allowbreak }\textbf{order\_\allowbreak of\_\allowbreak addition} & \texttt{string?} & The order in which components were added and any pre-incubation steps. \emph{Examples: Protein + compound pre-incubated 15 min, then tracer added; All components added simultaneously; Compound + tracer added to protein; Not specified} \\
\addlinespace[2pt]
\textbf{detection}$^{\dagger}$ & \texttt{object} & \textit{How fluorescence polarization or anisotropy was measured. FP/FA is a homogeneous assay with no separation step.} \\
\addlinespace[1pt]
\textcolor{gray!70}{detection.\allowbreak }\textbf{readout\_\allowbreak mode}$^{\dagger}$ & \texttt{string} & Whether the assay reports fluorescence polarization or fluorescence anisotropy. \emph{Allowed: Fluorescence Polarization (FP); Fluorescence Anisotropy (FA); Not specified} \\
\textcolor{gray!70}{detection.\allowbreak }\textbf{readout\_\allowbreak unit}$^{\dagger}$ & \texttt{string} & The unit used for the raw readout values. \emph{Allowed: mP (millipolarization); Anisotropy (r); Not specified} \\
\textcolor{gray!70}{detection.\allowbreak }\textbf{instrument}$^{\dagger}$ & \texttt{string} & The plate reader used to measure FP/FA. \emph{Examples: BMG PHERAstar FSX; Tecan Infinite M1000 Pro; PerkinElmer EnVision; Molecular Devices SpectraMax Paradigm; ...} \\
\textcolor{gray!70}{detection.\allowbreak }\textbf{g\_\allowbreak factor\_\allowbreak correction} & \texttt{string} & Whether the instrument G-factor was determined and applied to correct for optical bias between detection channels. \emph{Allowed: Yes; No; Not specified} \\
\addlinespace[2pt]
\textcolor{gray!70}{detection.\allowbreak }\textbf{assay\_\allowbreak window} & \texttt{object} & \textit{The dynamic range of the assay, i.e. the difference in mP or anisotropy between fully bound and free tracer.} \\
\addlinespace[1pt]
\textcolor{gray!70}{detection.\allowbreak assay\_\allowbreak window.\allowbreak }\textbf{bound\_\allowbreak signal} & \texttt{string | number?} & The FP/FA value for the fully bound tracer (high control). \emph{Examples: 220; 220 mP; 0.18 r} \\
\textcolor{gray!70}{detection.\allowbreak assay\_\allowbreak window.\allowbreak }\textbf{free\_\allowbreak signal} & \texttt{string | number?} & The FP/FA value for the free tracer (low control). \emph{Examples: 80; 80 mP; 0.05 r} \\
\textcolor{gray!70}{detection.\allowbreak assay\_\allowbreak window.\allowbreak }\textbf{delta} & \texttt{string | number?} & The difference between bound and free signal. \emph{Examples: 140; 140 mP; 0.13 r} \\
\addlinespace[2pt]
\textbf{data\_\allowbreak analysis}$^{\dagger}$ & \texttt{object} & \textit{Details about how the raw data was processed and modeled.} \\
\addlinespace[1pt]
\textcolor{gray!70}{data\_\allowbreak analysis.\allowbreak }\textbf{endpoint\_\allowbreak reported}$^{\dagger}$ & \texttt{string} & The primary binding parameter reported in the publication. \emph{Allowed: Ki; IC50; Kd; Percent inhibition; pKi; pIC50; EC50; Other} \\
\textcolor{gray!70}{data\_\allowbreak analysis.\allowbreak }\textbf{ic50\_\allowbreak to\_\allowbreak ki\_\allowbreak conversion}$^{\dagger}$ & \texttt{string?} & The method used to convert IC50 to Ki, if applicable. For FP/FA competition assays, Cheng-Prusoff requires the tracer Kd and tracer concentration. \emph{Allowed: Cheng-Prusoff equation; Nikolovska-Coleska equation; Exact solution; Not applicable; Not specified} \\
\textcolor{gray!70}{data\_\allowbreak analysis.\allowbreak }\textbf{fitting\_\allowbreak model}$^{\dagger}$ & \texttt{string} & The model used to fit competition curves. \emph{Allowed: One-site competition; Two-site competition; Four-parameter logistic (4PL); Three-parameter logistic (3PL, Hill slope fixed to 1); Dose-response (variable slope); Other} \\
\addlinespace[2pt]
\textcolor{gray!70}{data\_\allowbreak analysis.\allowbreak }\textbf{hill\_\allowbreak slope\_\allowbreak reported}$^{\dagger}$ & \texttt{object} & \textit{Whether a Hill slope (nH) was reported and its value, if available. nH $\neq$ 1 may indicate compound aggregation, cooperativity, or assay interference.} \\
\addlinespace[1pt]
\textcolor{gray!70}{data\_\allowbreak analysis.\allowbreak hill\_\allowbreak slope\_\allowbreak reported.\allowbreak }\textbf{reported}$^{\dagger}$ & \texttt{string} &  \emph{Allowed: Yes, fixed to 1; Yes, variable; Not reported} \\
\textcolor{gray!70}{data\_\allowbreak analysis.\allowbreak hill\_\allowbreak slope\_\allowbreak reported.\allowbreak }\textbf{value} & \texttt{number?} & The Hill slope value if reported as variable. Null if fixed or not reported. \\
\textcolor{gray!70}{data\_\allowbreak analysis.\allowbreak }\textbf{data\_\allowbreak normalization} & \texttt{string?} & How raw FP/FA data was normalized before curve fitting. \emph{Allowed: Percent inhibition relative to controls; Fraction bound; Raw mP values fitted directly; Raw anisotropy values fitted directly; Not specified} \\
\textcolor{gray!70}{data\_\allowbreak analysis.\allowbreak }\textbf{software}$^{\dagger}$ & \texttt{string} & The software used for curve fitting and data analysis. \emph{Examples: GraphPad Prism 9; GraphPad Prism 10; SigmaPlot; GeneData Screener; ...} \\
\addlinespace[2pt]
\textcolor{gray!70}{data\_\allowbreak analysis.\allowbreak }\textbf{replicates}$^{\dagger}$ & \texttt{object} & \textit{The number of replicates and independent experiments.} \\
\addlinespace[1pt]
\textcolor{gray!70}{data\_\allowbreak analysis.\allowbreak replicates.\allowbreak }\textbf{technical\_\allowbreak replicates}$^{\dagger}$ & \texttt{string} & Number of replicates per concentration within a single experiment. \emph{Examples: Duplicate; Triplicate; Singlicate; Not specified} \\
\textcolor{gray!70}{data\_\allowbreak analysis.\allowbreak replicates.\allowbreak }\textbf{independent\_\allowbreak experiments}$^{\dagger}$ & \texttt{string | number} & Number of independent experiments (n). Use a string to capture ranges. \emph{Examples: 3; 2-4; $\geq$3; Not specified} \\
\addlinespace[2pt]
\textbf{artifact\_\allowbreak controls} & \texttt{object} & \textit{Controls and checks for common FP/FA artifacts that can produce false positives or inaccurate affinities.} \\
\addlinespace[1pt]
\textcolor{gray!70}{artifact\_\allowbreak controls.\allowbreak }\textbf{compound\_\allowbreak fluorescence\_\allowbreak interference}$^{\dagger}$ & \texttt{string} & Whether the authors checked for intrinsic fluorescence of test compounds at the assay wavelengths, which can artifactually alter FP/FA readings. \emph{Allowed: Tested and excluded interfering compounds; Tested, no interference observed; Counter-screen performed; Not specified} \\
\textcolor{gray!70}{artifact\_\allowbreak controls.\allowbreak }\textbf{compound\_\allowbreak aggregation\_\allowbreak check} & \texttt{string} & Whether compound aggregation was assessed (e.g., dynamic light scattering, detergent sensitivity, Hill slope analysis). \emph{Allowed: Yes, detergent sensitivity tested; Yes, DLS performed; Yes, Hill slope monitored; No; Not specified} \\
\textcolor{gray!70}{artifact\_\allowbreak controls.\allowbreak }\textbf{inner\_\allowbreak filter\_\allowbreak effect} & \texttt{string} & Whether the authors checked for inner filter effect (absorption of excitation or emission light by the test compound at high concentrations), which can cause apparent FP changes unrelated to binding. \emph{Allowed: Tested; Not tested; Not specified} \\
\textcolor{gray!70}{artifact\_\allowbreak controls.\allowbreak }\textbf{orthogonal\_\allowbreak assay\_\allowbreak confirmation} & \texttt{string?} & Whether hits were confirmed by an orthogonal (non-FP) binding assay. \emph{Examples: ITC; SPR; TR-FRET; AlphaScreen; ...} \\
\end{longtable}
\end{small}

\section{Extraction Pipeline Implementation Details}
\label{app:implementation}

\begin{figure}[htb!]
\centering
\includegraphics[width=1\textwidth]{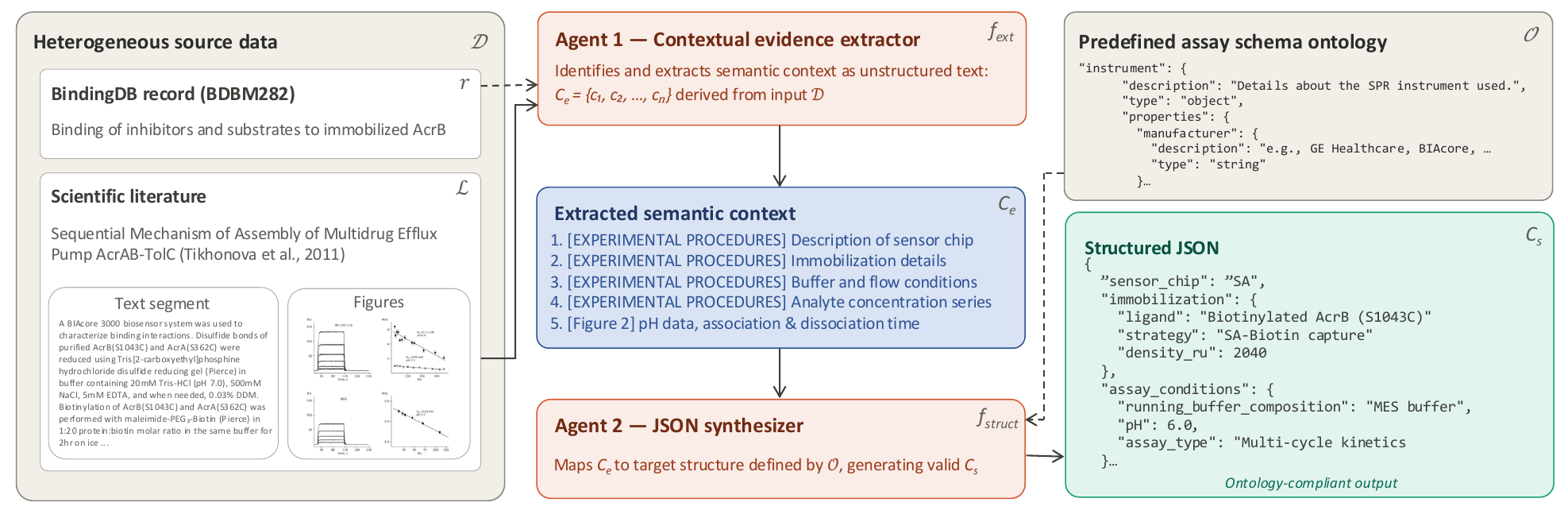}
\caption{Two-stage agentic extraction framework. Solid arrows denote data flow;
    dashed arrows denote conditioning inputs (prompts). The BindingDB
    record $r$ queries Agent~1 ($f_{\text{ext}}$), which extracts semantic
    context $\mathcal{C}_e$ from the associated literature $\mathcal{L}$.
    The assay schema ontology $\mathcal{O}$ conditions Agent~2
    ($f_{\text{struct}}$), which maps $\mathcal{C}_e$ to an
    ontology-compliant structured record $\mathcal{C}_s$.}
\label{fig:two_stage_agentic_extraction_framework}
\end{figure}

We describe the implementation of the assay-extraction pipeline used to
populate structured records from primary literature, using Surface Plasmon
Resonance (SPR) as the running example. The pipeline is decomposed into two
stages. \emph{Stage 1} (paragraph localization) operates on rendered page
images and is responsible for finding and copying the verbatim experimental
text from the paper, its supplementary information, and -- when needed --
referenced literature. \emph{Stage 2} (schema filling) consumes only the
text returned by Stage 1 and produces a structured JSON record conforming to
the SPR pre-defined schema. Decomposing the task in this way has two practical
consequences: (i) the expensive vision context (page images) is paid only
once per unique \texttt{(PMID, DESCRIPTION)} pair, while protein--ligand
pairs that share the same assay description reuse the cached result; and
(ii) Stage 2 reduces to text-only inference, so the same model can be served
without an image processor, or replaced with a smaller text-only LM with no
change to the rest of the pipeline.

\subsection{Models and Decoding}
\label{sec:models-decoding}

Both stages use \textsc{Qwen3.5-27B} (the VL variant for Stage 1, the same
checkpoint in text-only mode for Stage 2). Weights are loaded in
\texttt{bfloat16} with HuggingFace \texttt{device\_map="auto"} on a single
NVIDIA GPU. We disable the model's optional ``thinking'' channel
(\texttt{enable\_thinking=False}) so that the response is the raw assistant
turn, simplifying JSON parsing.

Decoding is fully deterministic: we set $\tau = 0$ and use greedy generation
(\texttt{do\_sample=False}) for both stages. The maximum number of newly
generated tokens is \texttt{max\_new\_tokens}\,=\,32,768, which we found
sufficient for both the verbatim paragraph copy in Stage 1 and the
structured JSON in Stage 2. Input page images are produced by
\texttt{fitz (1.24.14)}, one image per page, with no resolution downscaling. We do
not impose a hard cap on the number of pages for the main paper, but
supplementary PDFs longer than 50 pages are skipped (and the skip logged) to
prevent context-window overflows on outlier supplements.

\subsection{Prompt Templates}
\label{sec:prompts}

Each stage is driven by a single, schema-specific prompt; the SPR templates
are reproduced in condensed form below. We use plain JSON as the output
contract in both stages and post-process the response by stripping any
\verb|```json| code fences before parsing. Responses that fail
\texttt{json.loads} are treated as Stage failures and skipped; we do not
retry with a temperature-based sampler.

\paragraph{Stage 1 -- paragraph localization.} The vision model receives
the full set of page images followed by the prompt below. The
\texttt{<assay\_description>} placeholder is filled with the BindingDB
\texttt{DESCRIPTION} field for the current pair.

\begin{quote}\small\ttfamily
You are an expert scientific reader analyzing a research paper about
Surface Plasmon Resonance (SPR) experiments.\\[2pt]
Task: Find and extract the COMPLETE ORIGINAL text that describes the
following SPR assay experiment.\\[2pt]
Assay Description: <assay\_description>\\[2pt]
Instructions:
\begin{enumerate}
\item Search the paper for where this SPR assay is described (Methods
first, then Results, then figure captions).
\item Extract the COMPLETE ORIGINAL text -- methods paragraphs, numerical
values visible in sensorgrams, relevant figure captions, and tables.
\item Do NOT summarize or paraphrase.
\item If the paper refers to a previous publication for protocol details,
locate the exact numbered entry in the bibliography and copy the full
citation.
\end{enumerate}
Output format (JSON):\\
\{\,"original\_paragraph": \{"<key>": "<text>", ...\},\\
\phantom{\{}"reference\_number\_in\_text": "...",\\
\phantom{\{}"references\_previous": "..." \}
\end{quote}

The keys in \texttt{original\_paragraph} are intentionally free-form
location labels (e.g.\ ``Materials and Methods'', ``Figure~2 caption'',
``Supporting Information S1''). When the same description is also found in
the supplementary PDF or in a referenced paper, the agent merges the two
dictionaries with prefixed keys
(\texttt{[Supp <file>] <key>}, \texttt{[Ref PMID <id>] <key>}) so
provenance survives into Stage 2.

\paragraph{Stage 2 -- schema filling.} Stage 2 receives only the
concatenated text from Stage 1 (no images) and is asked to fill the SPR
pre-defined schema. The full template enumerates each field with a one-line type
description and a normalization rule; the \emph{prescriptive} parts of the
prompt are:

\begin{quote}\small\ttfamily
You are an expert scientific reader analyzing extracted text from a
research paper about Surface Plasmon Resonance (SPR) experiments.\\[2pt]
Task: Extract structured assay parameters from the following text.\\[2pt]
Assay Description: <assay\_description>\\
=== EXTRACTED TEXT FROM PAPER ===\\
<concatenated\_paragraphs>\\
=== END OF EXTRACTED TEXT ===\\[2pt]
Instructions:
\begin{enumerate}
\item Analyze the extracted text carefully to identify SPR assay parameters.
\item Extract structured assay parameters into the fields described below.
\item If a parameter is not mentioned in the text, use \texttt{null}.
\end{enumerate}
STRUCTURED PARAMETERS TO EXTRACT: \\
<pre-defined schema>
\end{quote}
The complete field-level schemas for all four assay types (SPR, ITC, FPA, RBA) are given in Appendix~\ref{app:schemas}.

\subsection{Reference Resolution}
\label{sec:reference-resolution}

A non-trivial fraction of method sections defer to earlier publications
(``SPR experiments were performed essentially as described in [27]'').
Without resolving such pointers, Stage~2 receives a paragraph that is
correct but uninformative and the structured record is mostly null. We
therefore add a reference-resolution step that is invoked (i)~after a
successful Stage~1 extraction whose \texttt{references\_previous} field
is non-empty, and (ii)~when Stage~1 fails to find any relevant paragraph
in the main paper or its supplementary. The same procedure is reused in
both cases; the only difference is whether the resolved paragraph is
\emph{combined} with or \emph{substituted} for the main-paper paragraph.
Recursion is capped at \texttt{max\_reference\_depth}\,=\,1 so that we
follow only direct references; transitive ``as described in~[X], which
is described in~[Y]'' chains are not chased, both because they are rare and because each additional hop
multiplies the failure surface.

The goal of the procedure is, starting from the model's
\texttt{references\_previous} field, to recover the \emph{PMID} of the
cited paper so that its PDF can be fetched from PubMed Central and
pushed back through Stage~1. To do this reliably, the Stage~1 prompt is
written to also report the in-text reference number (e.g.\ ``27''). When
such a number is present, we (i)~convert the source PDF to markdown
with MinerU~\citep{niu2025mineru2} and
(ii)~locate the matching bibliography entry by regular-expression match
on the two numbering conventions dominant in chemistry/biology
journals, \verb|(N) Author, ...| and \verb|N. Author, ...| The recovered
citation string (e.g.\ ``Smith, J.\ et al.\ \emph{J.\ Med.\ Chem.}\
2015, 58, 1234'') is then resolved to a PMID via the NCBI E-utilities
\texttt{esearch} endpoint, in three substrategies tried in order: a DOI
match if the citation contains one, a title match (the title is
extracted by a one-shot LM query with a regex fallback), and finally a
structured author$+$year$+$journal$+$volume$+$page query. The first
substrategy that returns a unique hit is accepted.

We deliberately read the bibliography from MinerU's text-layer
extraction rather than from the rasterized page images used in
Stage~1: bibliography entries are typeset in small font and densely
packed with easily-confused tokens (author initials, volume/page
numbers, DOIs), where the vision encoder's effective resolution
produces frequent OCR-style errors. Because the citation string
recovered this way is verbatim from the paper, this design essentially
eliminates citation hallucination; the only remaining failure mode is
the PubMed lookup itself, and any reference that we cannot resolve to
a PMID is appended (with the offending citation string) to a per-run
\texttt{chase\_log} for auditability.

Once a referenced PMID is resolved, the corresponding PDF is fetched
from PubMed Central, pushed through Stage~1 with the same prompt, and
its \texttt{original\_paragraph} dictionary is merged into the
main-paper one with a provenance prefix \verb|[Ref PMID <id>] <key>|
(see Section~\ref{sec:prompts}), so Stage~2 can read both the original
and the cited methods text in a single context. At most three
referenced PMIDs are followed per paper.

\section{Compute Resources}
\label{app:compute-resources}
All experiments reported in this paper were performed on a single shared compute node equipped with two Intel Xeon Gold 6442Y CPUs (48 physical cores in total), 1 TB of system memory, and eight NVIDIA H100 80GB HBM3 GPUs (CUDA 12.6, driver 560.35.05). Individual workloads were assigned to one or more GPUs; no single experiment required more than eight GPUs concurrently.



\end{document}